\documentclass[aps,prb,reprint,floatfix,superscriptaddress]{revtex4-1}
\usepackage[english]{babel}
\usepackage[T1]{fontenc}
\usepackage{bbold}
\usepackage{epsfig}
\usepackage{amsmath}
\usepackage{hyphenat}
\usepackage{hyperref}
\usepackage{placeins}
\usepackage[justification=raggedright]{caption}
\usepackage{subcaption}	

\usepackage[usenames,dvipsnames]{color}

\usepackage{array}
\newcolumntype{C}[1]{>{\centering\arraybackslash}m{#1}}
\usepackage{overpic}

\begin{document}

\title{Disorder in tilted Weyl semimetals from a renormalization group perspective}
\author{Tycho S. Sikkenk}
\affiliation{Institute for Theoretical Physics and Center for Extreme Matter and Emergent Phenomena, Utrecht University, Leuvenlaan 4, 3584 CE Utrecht, The Netherlands}
\author{Lars Fritz}
\affiliation{Institute for Theoretical Physics and Center for Extreme Matter and Emergent Phenomena, Utrecht University, Leuvenlaan 4, 3584 CE Utrecht, The Netherlands}

\begin{abstract}
Isolated Weyl cones in a disordered environment do not show the phenomenon of Anderson localization due to the abscence of backscattering processes. However, besides the conventional three dimensional diffusive metal, an additional semimetallic phase can form. In this paper we study the effect of tilt on the physics of disorder in isolated Weyl cones. Our main finding is that while the semimetallic phase remains a stable phase, tilt decreases the size of the semimetallic region. Conversely, disorder increases the effective tilt of the quasiparticle excitations. 
\end{abstract}

\maketitle

\section{Introduction}

In condensed matter theory, the Dirac equation describes the behavior of low energy quasiparticles near the touching points of a set of twofold degenerate bands in the three dimensional dispersion of a class of materials known as Dirac semimetals (DSMs). These Dirac nodes are topologically protected by time reversal and inversion symmetry and cannot be removed without opening a gap in the spectrum. In Weyl semimetals (WSMs), at least one of these symmetries is broken and the Dirac nodal points disassociate into two Weyl nodes that are separated in momentum space. There is a non-zero flux of Berry curvature between the nodes that enriches them with an opposite monopole charge, or chirality, by Gauss' law. Band structure calculations predicted WSMs in several material families \cite{weng2015weyl,huang2015weyl} and Weyl fermion states have been discovered experimentally in materials such as TaAs, NbAs, TaP and NbP \cite{Lv2015Experimental,xu2015discovery,xu2015discovery2,yang2015weyl,lv2015observation,xu2016observation,shekhar2015extremely}. Because of the unique features of their nodes, WSMs can exhibit a range of fascinating properties not observed in graphene, DSMs or other topological semimetallic systems. On their surface they harbor open Fermi surfaces connecting the node projections, implying novel contributions to various transport properties \cite{Shi2017, Baum2015}. Such surface arcs have also observed in angle-resolved photoemission spectroscopy (ARPES) experiments carried out for these systems.

The topological properties of these systems also lead to a remarkable robustness against weak perturbations, including disorder from dilute impurities, \cite{bera2016dirty}, since the Weyl nodes can only be annihilated by joining pairs of opposite chirality \cite{wan2011topological}. This has important consequences for the properties under disorder, most notably of which Anderson localization~\cite{Anderson1958,Abrahams1979}. For three dimensional metals it is well established that generically a mobility edge separating localized states from extended states develops as a function of disorder. The associated phase transition connects the diffusive metal (DM) with states above the mobility edge to the Anderson insulator on the other side. In Weyl systems the situation can be different in two ways: (i) A single Weyl node, very similarly to the surface Dirac fermion of a three dimensional strong topological insulator, cannot be localized by disorder because no backscattering can occur~\cite{Bardarson2007}. In reality, however, every Weyl node in a Weyl semimetal must be accompanied by its chiral partner, which then allows for backscattering through processes connecting the two cones. If the disorder is sufficiently long-range backscattering can be suppressed by the distance between the nodes in reciprocal space~\cite{Altland2016}. Indeed, intercone scattering has typically been found to have a much smaller amplitude than intracone scattering in realistic samples \cite{Xiong2015}. In the following we concentrate on a single Weyl node, assuming intercone processes are negligible compared to scattering within the cone. (ii) For a single Weyl node it was pointed out that there can be an additional phase at weak disorder that is semimetallic (SM)~\cite{Fradkin1986,Ominato2014,Sbierski2014,Syzranov2015,Pixley2015,Pixley2016A}. This point of view is currently challenged by numerical simulations which indicate the presence of rare regions~\cite{Pixley2016}. However, for intermediate energy scales the critical point between the semimetal and diffusive metal should still control the physics, which is the point of view taken in this paper.

More recently, it was appreciated that Weyl cones in condensed matter systems, in contrast to fundamental Weyl fermions, do not have to be perfectly isotropic. The Femi velocity may take unequal values in different directions. Such anisotropy is common in Weyl semimetals and has significant consequences for various observable quantities \cite{Rodionov2015,Trescher2015}. Alternatively, the Weyl spectrum may continuously be tilted away from its upright position breaking time reversal symmetry. Recently, it was first predicted \cite{soluyanov2015typeii,Bergholtz2015} and later experimentally confirmed \cite{Koepernik2016, Huang2016,Belopolski2016,Zyuzin2016} that the conical dispersion can even be tilted over. These systems, commonly referred to as type-II Weyl semimetals, develop a Fermi surface in which electron and hole pockets coexist with the Weyl touching points. This leaves clear signatures in their thermodynamic and transport properties, e.g. magnetic breakdown at nodal energy, unusual magneto-respsonse and the emergence of an intrinsic anomalous Hall effect \cite{OBrien2016, Yu2016,Zyuzin2016}. Since nodal quasiparticles can scatter resonantly to electron or hole surfaces, inclusion of disorder leads to considerably more complicated situations.

In contrast, subcritical tilts qualitatively preserve the structure of the pointlike Fermi surface. Nevertheless, they can have a definite influence on the transport properties of type-I Weyl cones. For instance, the conductance and Fano factor show strongly anisotropic responses upon increasing tilt \cite{Trescher2015}. Tilted Weyl cones support an increased amount of states at energies away from the nodal point and so enhance the effects of level broadening under disorder.

In this paper we investigate the interplay of finite subcritical tilts and potential disorder within the framework of the renormalization group. Our findings confirm earlier conclusions based on numerical methods and self-consistent Born approximation (SCBA) analysis \cite{Trescher2017}, and additionally provide new analytic insights to describe disordered tilted Weyl cones.  

The remainder of this work is organized as follows: Sec.~\ref{sec:model} introduces the model. In Sec.~\ref{sec:rg} we study this model using an RG approach and discuss the resulting flow. Our main results consist of a set of analytical equations describing the scaling of the density of states when taking disorder effects into account. We show also that finite tilts lower the disorder strength of the SM-DM phase transition. Disorder generically renders the system anisotropic and furthermore increases the observable tilt, thereby enhancing the density of states. We finish with a conclusion in Sec.~\ref{sec:conc}.

\section{Model Setting}\label{sec:model}

The starting point of our investigations is given by a single titled Weyl cone described by the Bloch Hamiltonian
\begin{align}\label{eq:ham}
\mathcal{H}_\chi ({ \bf k}) =v \left(t \, \sigma_0 \, {\bf d} + \chi \, {\pmb \sigma} \right) \cdot { \bf k}
\end{align}
where $\sigma_0$ is the identity matrix and $\{ \sigma_x, \sigma_y, \sigma_z \}$ are the standard Pauli matrices. The Fermi velocity is given by $v$, while the strength and direction of the tilt are parametrized by  $t$ and ${\bf d}$ ($|{\bf d }|=1$), respectively. The chirality of the cone is set by $\chi = \pm 1$. The eigenspectrum of Eq.~\eqref{eq:ham} reads
\begin{align} \label{eq:tilt_disp}
E_s ({\bf k})=v \left(t \, {\bf d} \cdot {\bf k}+ s \,  k \right) = v \left( t k_\parallel  + s \sqrt{k_\parallel^2 + k_\perp^2} \right) \;, 
\end{align}
where $s = \pm 1$ distinguishes electron and hole bands. We furthermore introduce a momentum parametrization where $k_\parallel$ is oriented in the direction of ${\bf d}$ while $k_\perp$ is a radial coordinate in the plane perpendicular to ${\bf{d}}$. Upon increasing $t$ the band structure progressively tilts over until the cone acquires one dispersionless direction at $t=1$ thereby effectively becoming metallic (in principle higher order in momentum terms have to be included to stabilize the theory). Concentrating on values $|t|<1$ the density of states (DoS) behaves according to
\begin{eqnarray} \label{eq:DoS}
\rho_0^t (\omega) = 
\frac{\omega^2}{2\pi^2 (1-t^2)^2 v^3} \;
\end{eqnarray}
which becomes singular for $|t| \to 1$, consistent with the system going metallic. 

We consider quenched scalar chemical potential disorder obeying a Gaussian white-noise distribution with zero mean \cite{Goswami2011,Altland2016,Roy2016},
\begin{align} \label{eq:disorder_corr}
\langle V( {\bf x} ) \rangle = 0 \; , \qquad \langle V({\bf x}) V({\bf x}') \rangle = \Gamma^2 \, \delta^3 ({\bf x} - {\bf x}')\;.
\end{align}
We are interested in average properties of the disordered system. Therefore, we promote the disorder potential to a field which is integrated over corresponding to an averaged partition function
\begin{align}
\bar{Z}	 	&=  \int \mathcal{D} [\psi_\chi^\dagger, \psi_\chi] \, \mathcal{D} V \, \exp \left\{ - \bar{S} [\psi_\chi^\dagger,\psi_\chi; V] \right\}, \\
\bar{S}_\chi	&= S_V + S_{0;\chi}^{t,\lambda} +  S_{\Gamma;\chi} + S_{\bar{\Gamma};\chi}\;. \label{eq:action}
\end{align}
Here, we have rescaled $V$ by $\Gamma$ in both the disorder coupling term and the Gaussian weight introduced during the averaging procedure to obtain the expressions
\begin{align}
S_V 			&= \frac{1}{2} \int \mathrm{d} {\bf x} \;  V^2 ( {\bf x})\;, 		\label{eq:disorder_action} \\
S_{\Gamma;\chi} 	&= \Gamma \int \mathrm{d} {\bf x} \, \mathrm{d} \tau \;  V( {\bf x}) \left(\psi_\chi^\dagger \sigma_0 \psi_\chi \right)\;. \label{eq:sigma0_vertex_action}
\end{align}
Philosophically, our procedure is equivalent to replicating the system and integrating out disorder, which leads to a retarded interaction in replica space. In contrast, we integrate out disorder only at every order in perturbation theory instead of exactly which provides us with technical advantages.

\begin{figure}[h]
\centering
  \begin{subfigure}[t]{0.46\textwidth}
  \centering
    \includegraphics[width=0.45\textwidth]{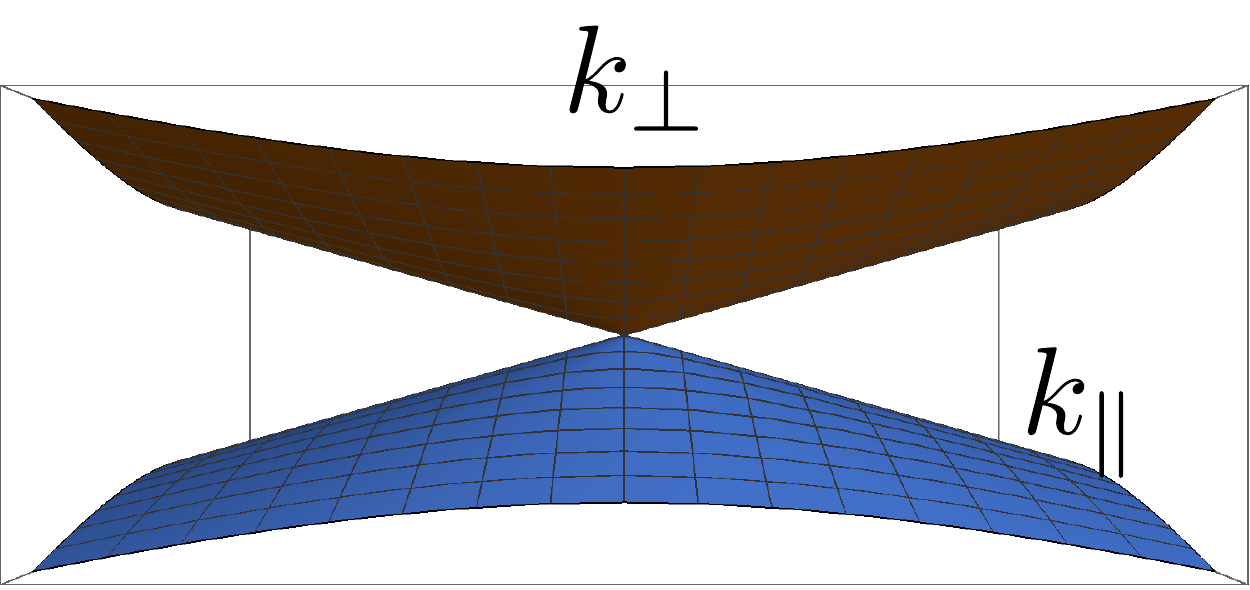}
    \includegraphics[width=0.45\textwidth]{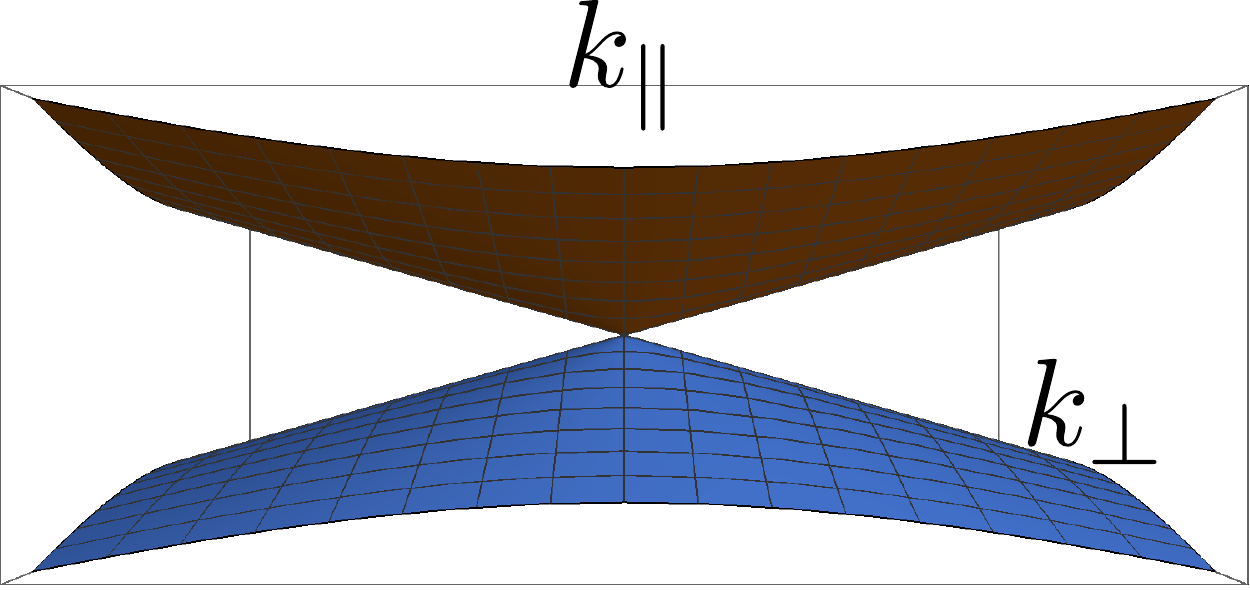}
    \caption{$v=1$, $t=\lambda=0$.}
    \label{fig:Dispersion_untilted}
  \end{subfigure}
  \qquad
  \begin{subfigure}[t]{0.46\textwidth}
  \centering
    \includegraphics[width=0.45\textwidth]{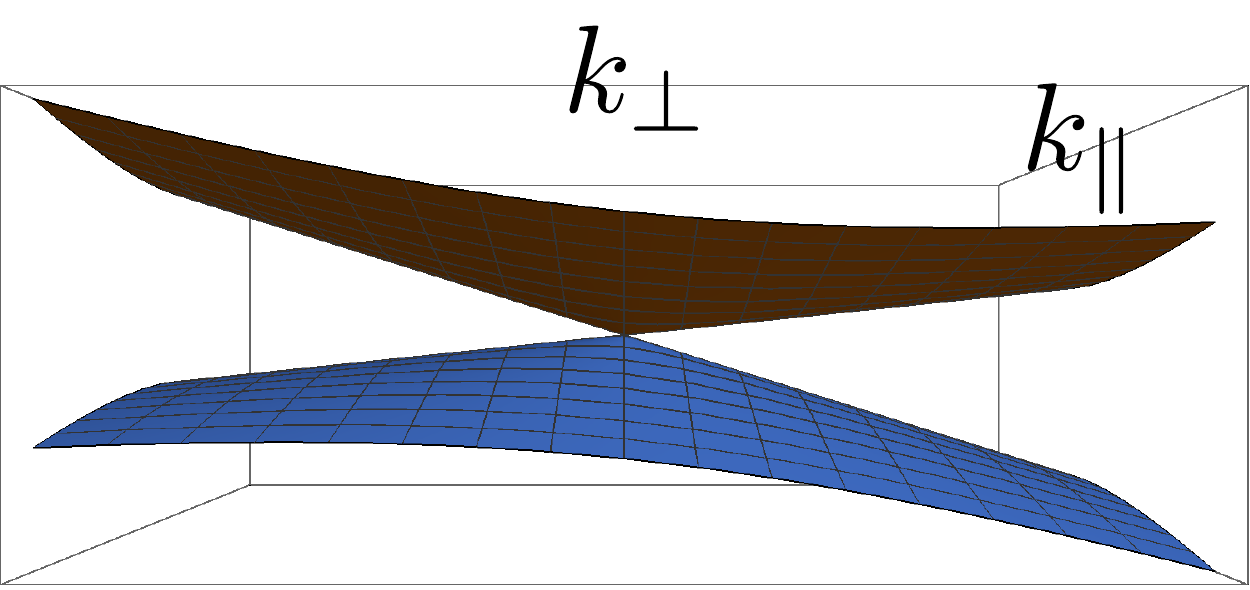}
    \includegraphics[width=0.45\textwidth]{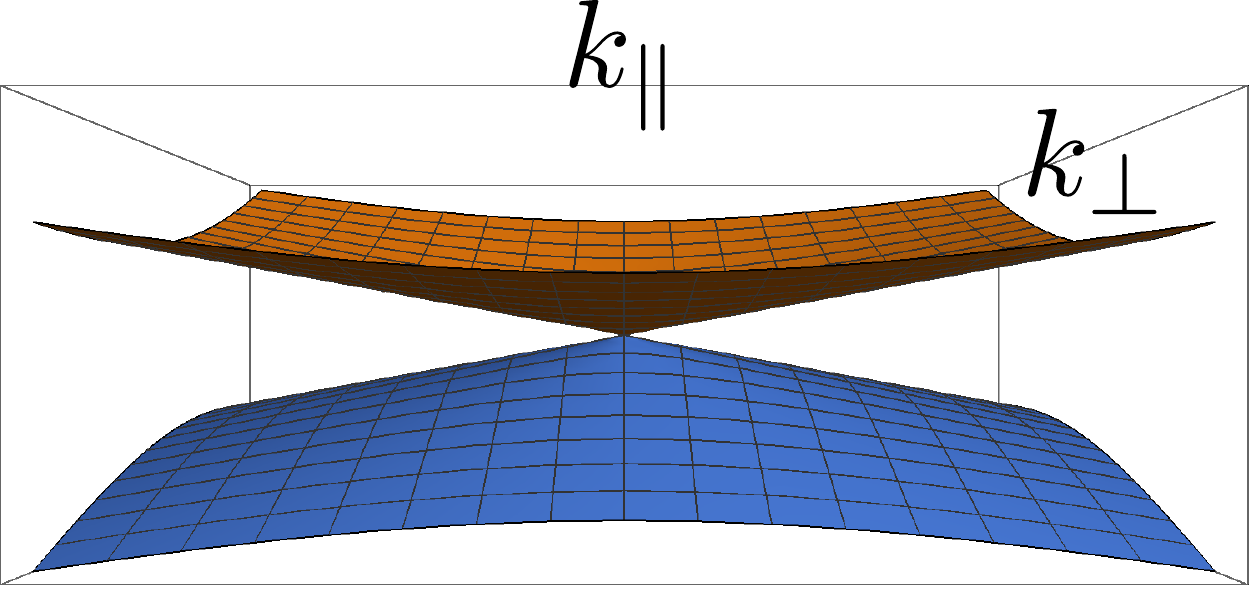}
    \caption{$v=1$, $t=0.5$, $\lambda=0$.}
    \label{fig:Dispersion_tilted}
  \end{subfigure}
  \qquad
  \begin{subfigure}[t]{0.46\textwidth}
  \centering
    \includegraphics[width=0.45\textwidth]{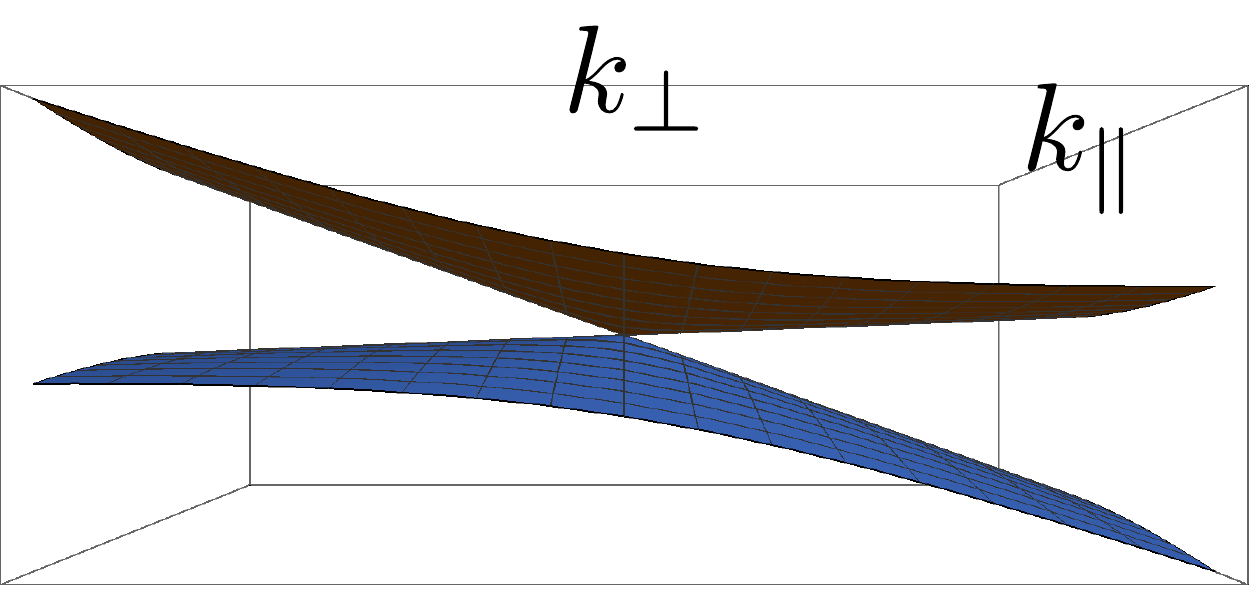}
    \includegraphics[width=0.45\textwidth]{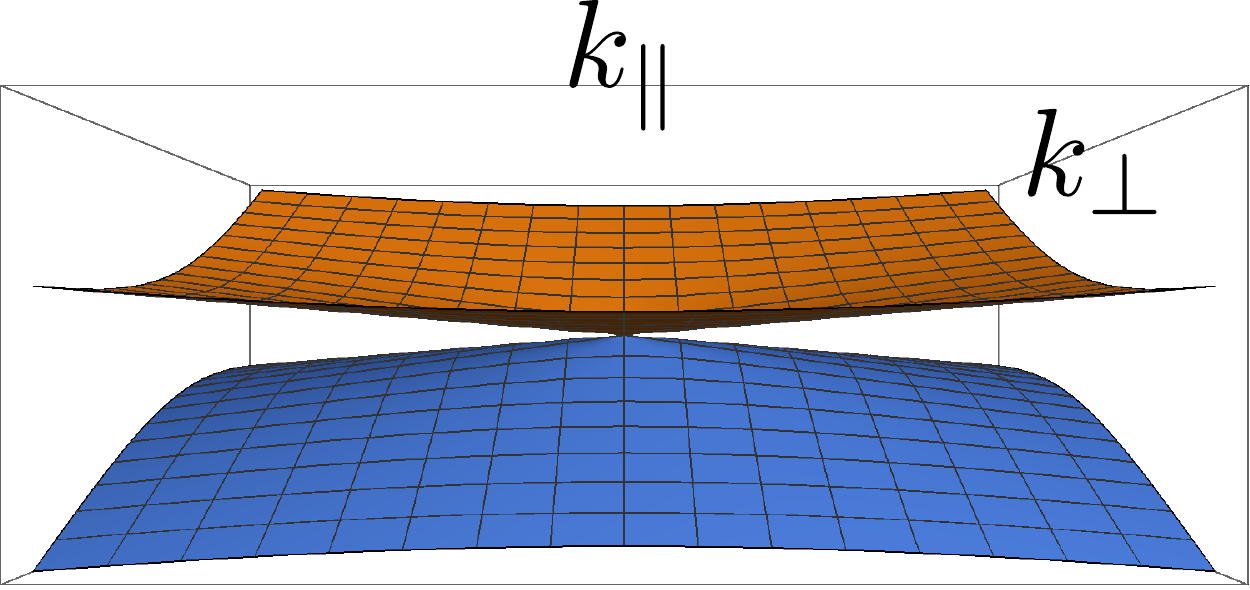}
    \caption{$v=1$, $t=\lambda=0.5$.}
    \label{fig:Dispersion_enhanced_tilt}
  \end{subfigure}
  \caption{The modified Weyl cone dispersion $E^{t,\lambda}$ for (a) the untitled case, (b) ordinary tilt and (c) modified tilting.}
  \label{fig:Modified_dispersion}
\end{figure}

The combination of tilt and disorder (Eq.~\eqref{eq:action}, Eq.~\eqref{eq:disorder_action}, and Eq.~\eqref{eq:sigma0_vertex_action}) is not closed under a one-loop RG transformation. The self-energy generates a marginal term $\sim i \omega \, \chi \, {\bf d} \cdot {\pmb \sigma}$, which we thus add to the original action. The tree level Weyl fermion action including this term is given by
\begin{align} \label{eq:fermion_action}
S_{0;\chi}^{t,\lambda} 	&= \int \mathrm{d} {\bf x} \, \mathrm{d} \tau \; \psi_\chi^\dagger \left[ ( \sigma_0 - \chi \lambda \, {\bf d} \cdot {\pmb \sigma} ) \, \partial_\tau \right. \\ 
				&\qquad \quad - \left. i v (t \, \sigma_0 \,  {\bf d} + \chi {\pmb \sigma} )\cdot {\pmb \partial} \right] \psi_\chi. \nonumber 
\end{align}
Further analysis shows that there is also an additional vertex that is generated under the renormalization group transformation. As a result we add a second type of disorder vertex:
\begin{align} \label{eq:chiral_vertex_action}
S_{\bar{\Gamma};\chi} = \bar{\Gamma} \int  \mathrm{d} {\bf x} \, \mathrm{d} \tau \; V( {\bf x}) \left(\psi_\chi^\dagger (- \chi {\bf d} \cdot {\pmb \sigma}) \psi_\chi \right)\;,
\end{align}
which obeys the same distribution function as the original vertex.
The altered free fermion Green function $G_{0;\chi}^{t,\lambda} (i \omega, {\bf k}) =  \left[ (i \omega + v t \, {\bf d} \cdot {\bf k}) \sigma_0 + \chi(v \, {\bf k} - i \omega \lambda \, {\bf d}) \cdot {\pmb \sigma} \right]^{-1}$ which results from Eq.~\eqref{eq:fermion_action} has excitations at modified energy-momentum pairs. The poles of the inverse propagator, determined from $\det [ G_{0;\chi}^{t,\lambda} ]=0$, yield the dispersion
\begin{align} \label{eq:enhanced_tilt_disp}
E_s^{t,\lambda} ({\bf k}) &= v^\lambda \left[ (t + \lambda) k_\parallel + s \sqrt{ k_\parallel^2 (1+ t \lambda)^2 + k_\perp^2 (1-\lambda^2)} \right],
\end{align}
where $v^\lambda = v / (1-\lambda^2)$. We plot excitation spectrum, Eq.(\ref{eq:enhanced_tilt_disp}), in Fig.(\ref{fig:Modified_dispersion}) for some parameter values. Comparison with Eq.(\ref{eq:tilt_disp}) teaches that the effect of $\lambda$ is twofold:
\begin{itemize}
\item[(1).] It introduces anisotropy into the system, inducing respective effective Fermi velocities 
\begin{align}
\quad \; v_\perp = v^\lambda \, \sqrt{1-\lambda^2} \; , \quad \; v_\parallel = v^\lambda \, (1 + t \lambda) \;,
\end{align}
perpendicular and parallel to the orientation of the tilt.
\item[(2).] It introduces an effective given by
\begin{align} \label{eq:eff_tilt}
t_\text{eff} = (t+ \lambda) / (1 + t \lambda)\;.
\end{align}
\end{itemize}
The presence of $\lambda$ is reflected in the density of states, $\rho_0^{t,\lambda} (\omega) = -\frac{1}{\pi} \int_{\bf k} \; \text{Im} \; \text{Tr} \; G_{0;\chi}^{t,\lambda} (\omega + i 0^+, {\bf k})$, according to 
\begin{align} \label{eq:DoS_obs}
\rho_0^{t,\lambda} (\omega) =  \omega^2 \; \, \frac{ (1 +t \lambda)^2}{ 2\pi^2 (1-t^2)^2 v^3} = \rho_0^{t_\text{obs}} (\omega).
\end{align}
Although increasing the anomalous tilt above unity implies tipping over the conical dispersion, Eq.~\eqref{eq:eff_tilt}, a vanishing quasiparticle weight ensures that Eq.~\eqref{eq:DoS_obs} remains completely regular for all values of $\lambda$. For observable quantities, the measurable tilt is given by
\begin{align} \label{eq:obs_tilt}
t_\text{obs} = \sqrt{ \frac{t \, ( t + \lambda)}{1 + t \lambda} } = \sqrt{t \, t_\text{eff}}.
\end{align}

\section{Renormalization Group Analysis}\label{sec:rg}

The tree level actions Eq.~\eqref{eq:disorder_action} and Eq.~\eqref{eq:fermion_action} remain invariant under anisotropic space-time rescaling $\tau \to e^{z l} \tau$, ${\bf x} \to e^l {\bf x}$ if $[\psi] = [V] = d/2$ for the fermion and boson field, $[v]= z-1$ for the Fermi velocity and $[\Gamma] = [\bar{\Gamma} ] = z - d/2$ for the disorder couplings. The remaining parameters are scale-invariant irrespective of the number of spatial dimensions, $[t]=[\lambda]=0$. Therefore, for three-dimensional Weyl semimetals ($d=3$, $z=1$) both disorder vertices are irrelevant at tree level. This opens up the possibility of the existence of a nontrivial quantum critical point separating a semimetallic weak disorder phase from the diffusive metal phase, as was extensively discussed in the untilted case \cite{Fradkin1986,Ominato2014,Sbierski2014,Syzranov2015}. 

\begin{figure}[h]
\centering
  \begin{subfigure}[b]{0.15\textwidth}
  \centering
    \includegraphics[width=0.9\textwidth]{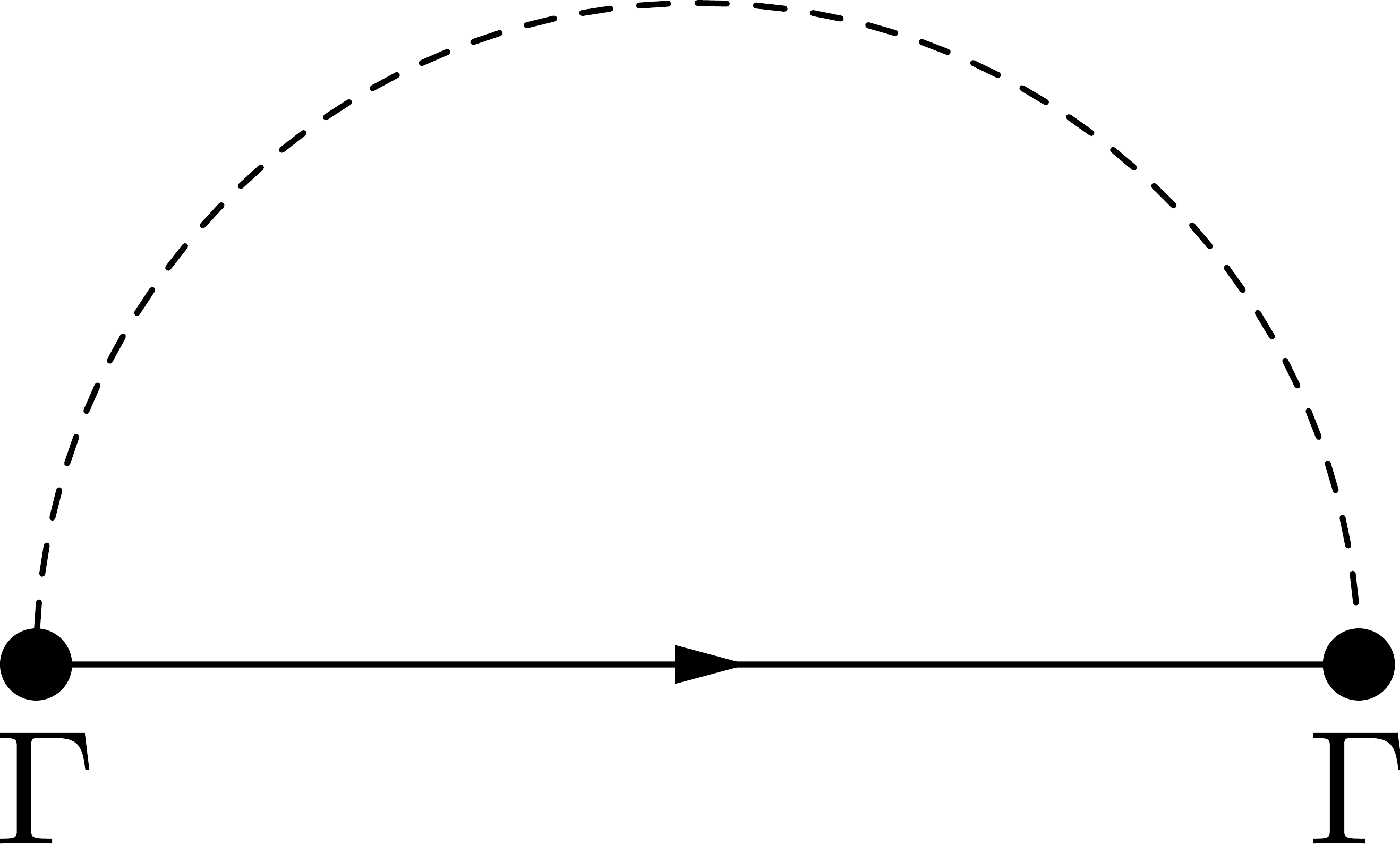}
    \caption{}
    \label{fig:SELF}
  \end{subfigure}
  \;
  \begin{subfigure}[b]{0.15\textwidth}
  \centering
    \includegraphics[width=0.9\textwidth]{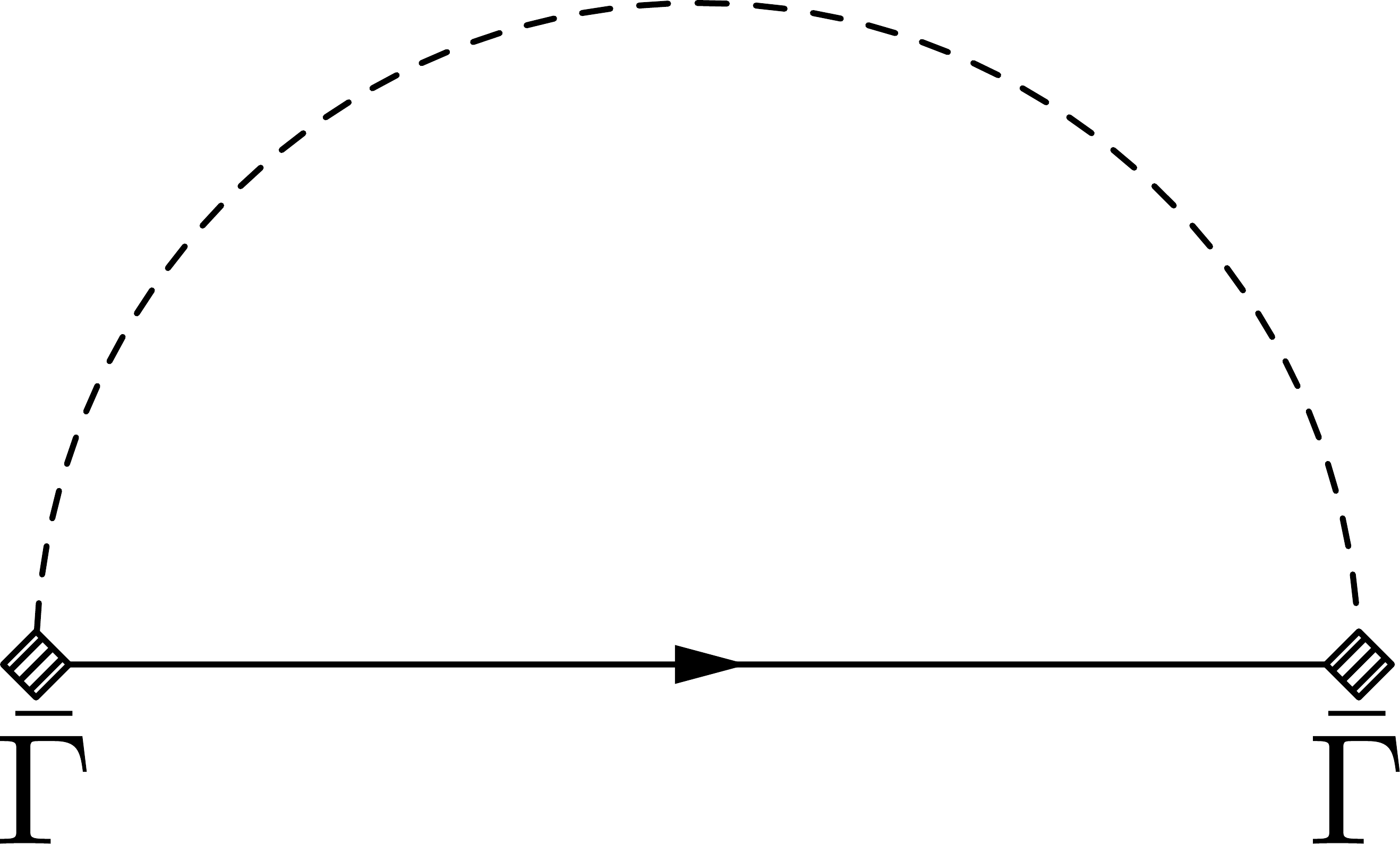}
    \caption{}
    \label{fig:SELF_VAR}
  \end{subfigure}
  \;
  \begin{subfigure}[b]{0.15\textwidth}
  \centering
    \includegraphics[width=0.9\textwidth]{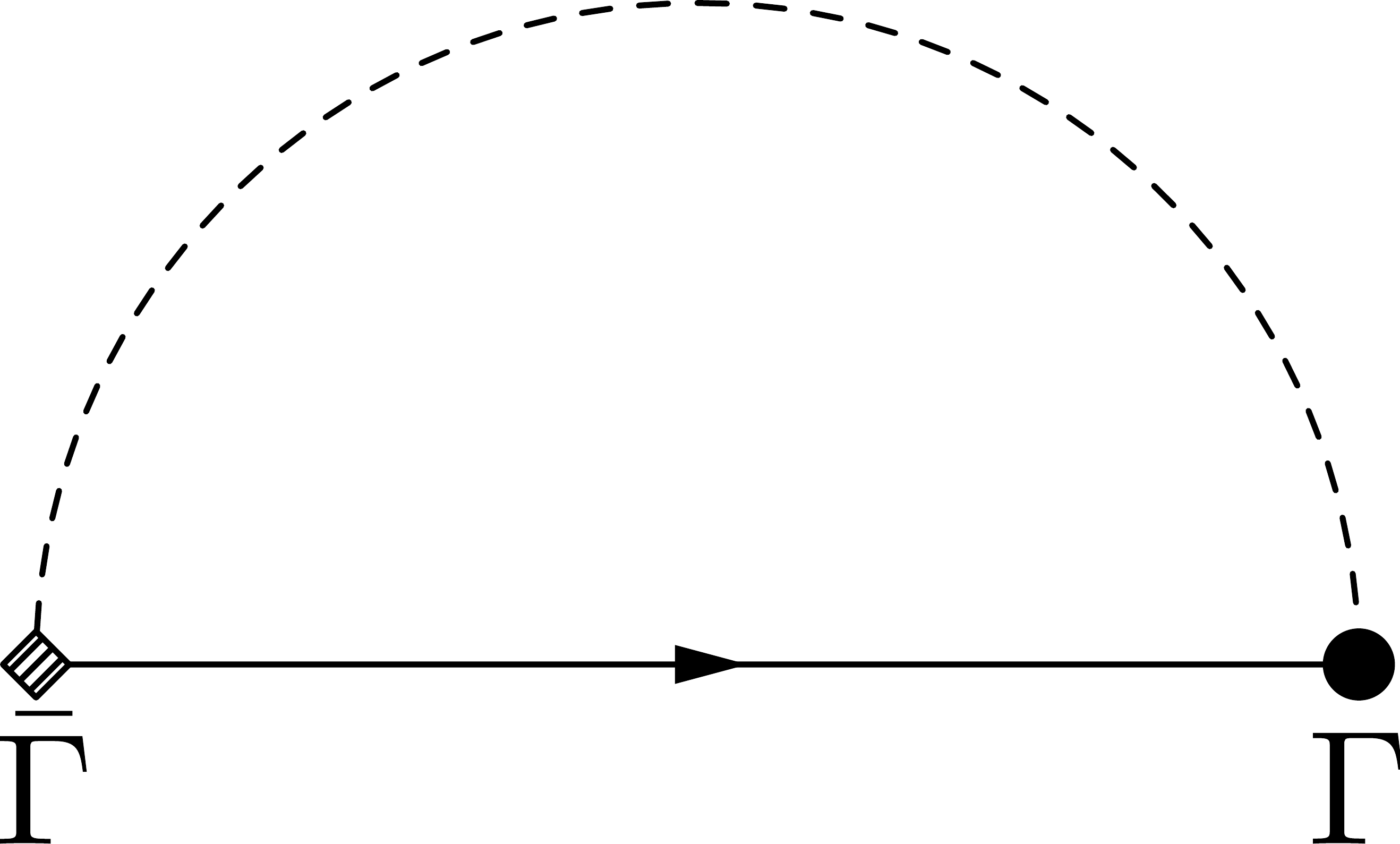}
    \caption{}
    \label{fig:SELF_COMBI}
  \end{subfigure}
   \caption{Diagrams contributing to the self-energy correction $\Sigma_\chi^{t,\lambda}$ to the modified naked Green function $G_{0;\chi}^{t,\lambda}$.}
   \label{fig:SELFS}
\end{figure}

\begin{figure}[h] 
\centering
  \begin{subfigure}[b]{0.15\textwidth}
  \centering
    \includegraphics[width=0.9\textwidth]{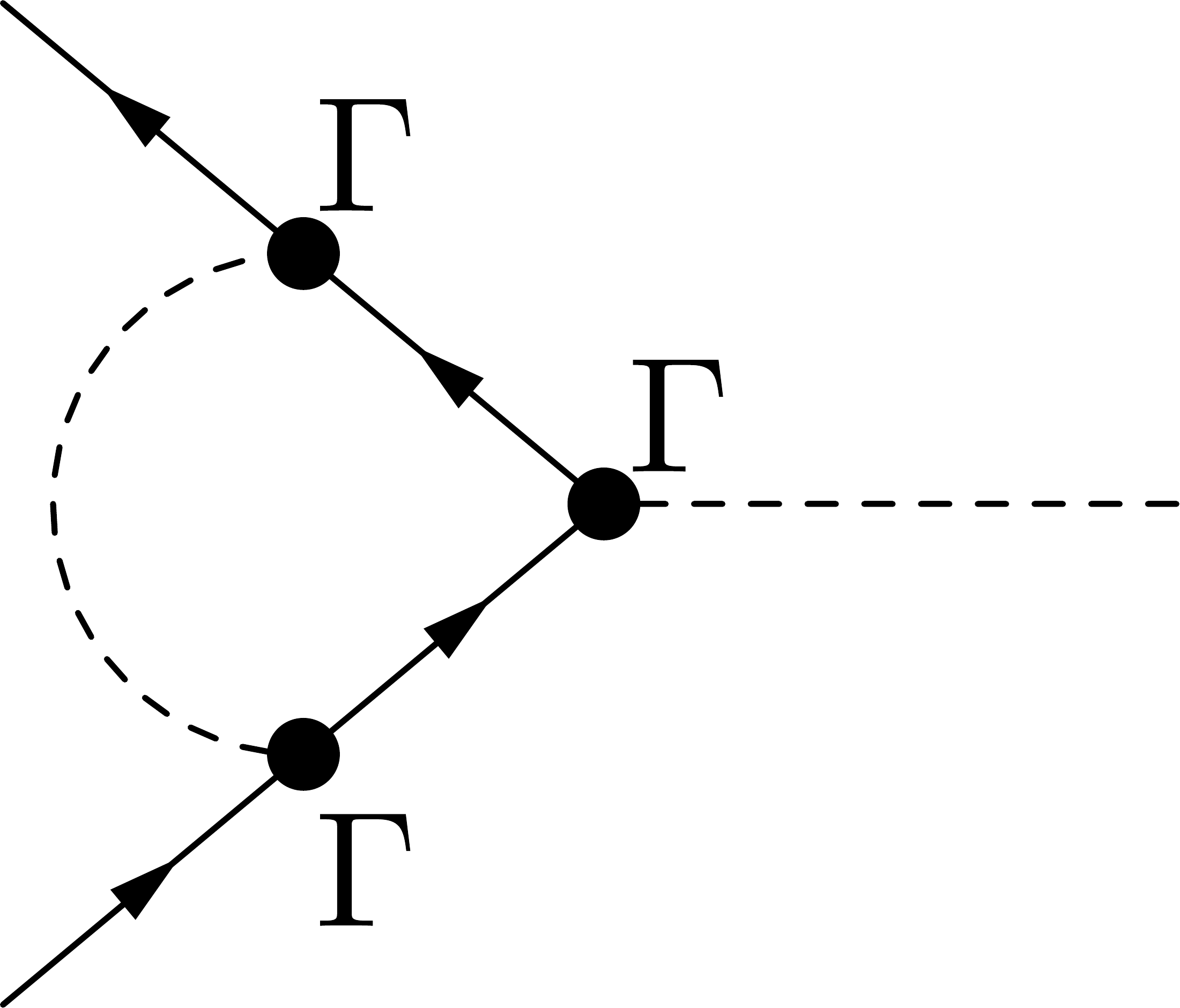}
    \caption{}
    \label{fig:VERTEX_CORR}
  \end{subfigure}
  \;
  \begin{subfigure}[b]{0.15\textwidth}
  \centering
    \includegraphics[width=0.9\textwidth]{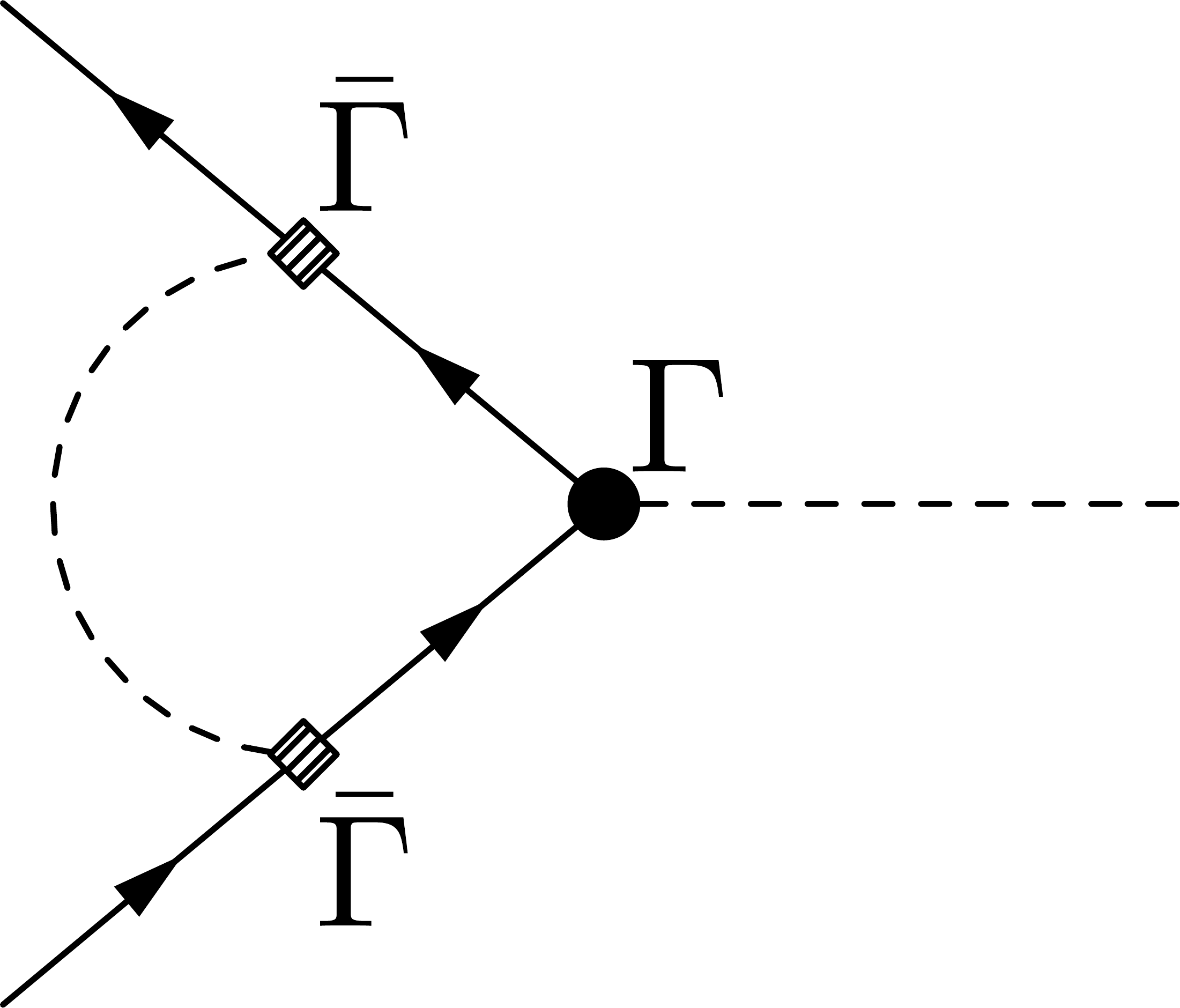}
    \caption{}
    \label{fig:VERTEX_CORR_COMBI1}
  \end{subfigure}
  \;
  \begin{subfigure}[b]{0.15\textwidth}
  \centering
    \includegraphics[width=0.9\textwidth]{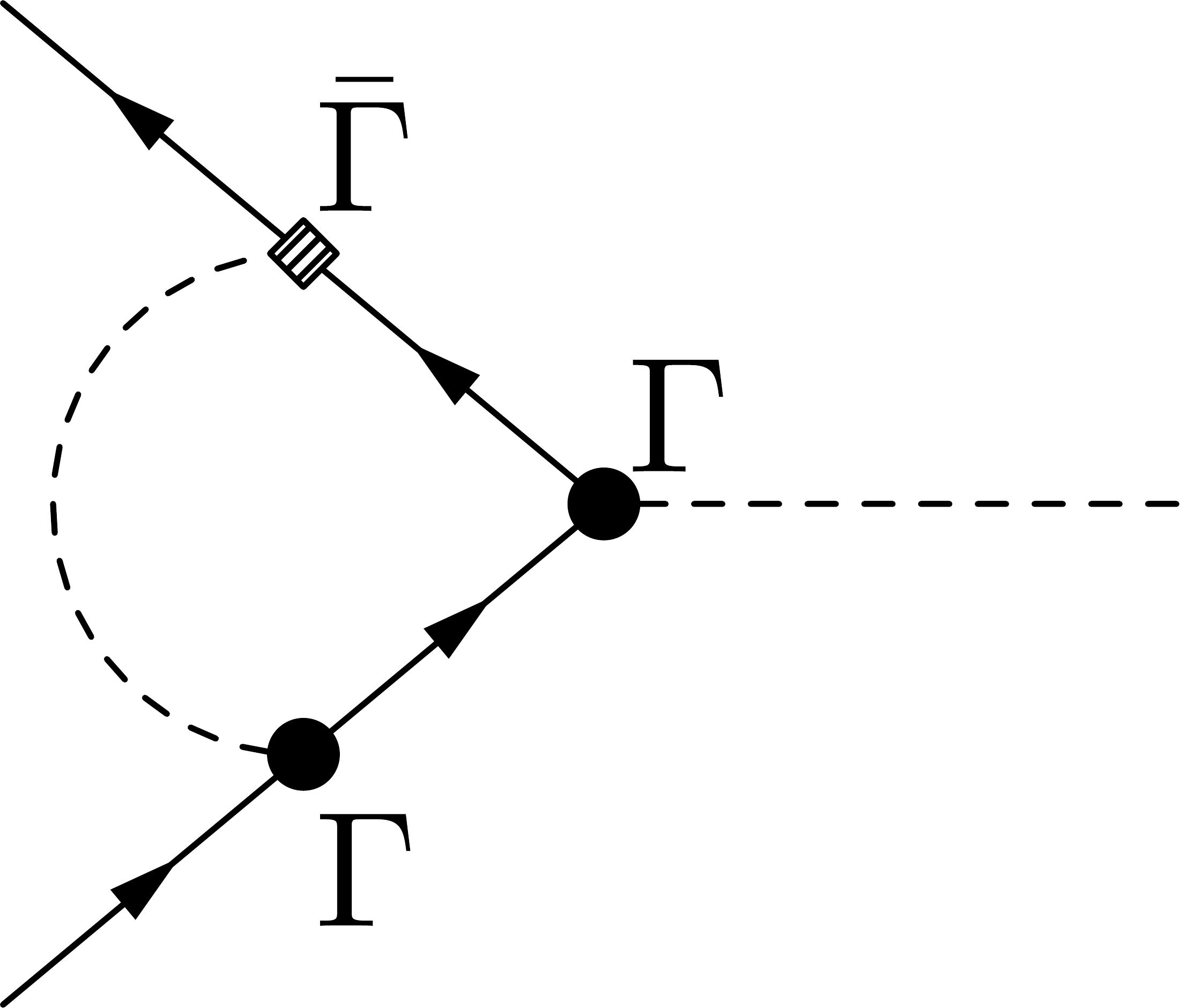}
    \caption{}
    \label{fig:VERTEX_CORR_COMBI2}
    \end{subfigure}
  \begin{subfigure}[b]{0.15\textwidth}
  \centering
    \includegraphics[width=0.9\textwidth]{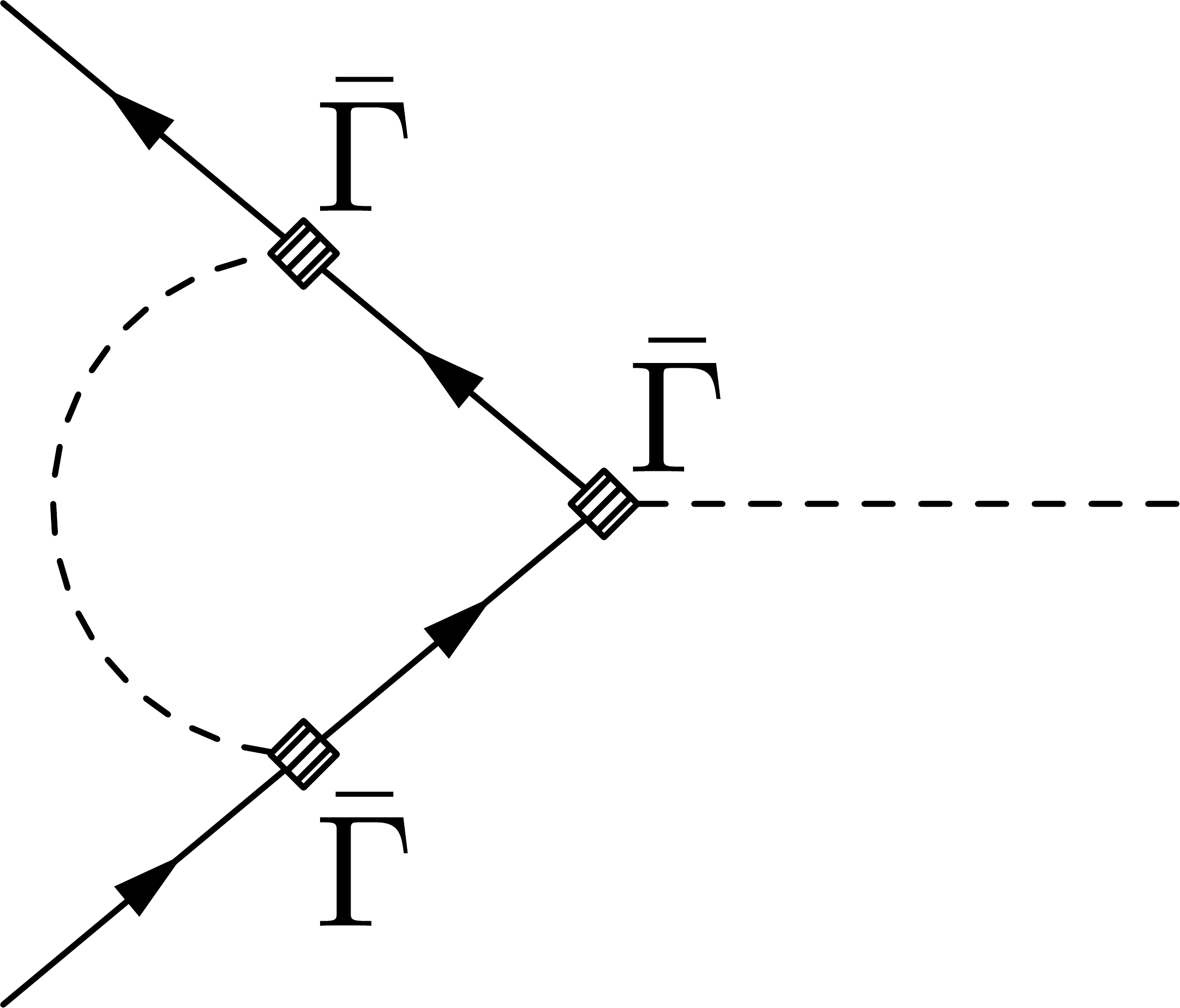}
    \caption{}
    \label{fig:VERTEX_CORR_VAR}
  \end{subfigure}
  \;
  \begin{subfigure}[b]{0.15\textwidth}
  \centering
    \includegraphics[width=0.9\textwidth]{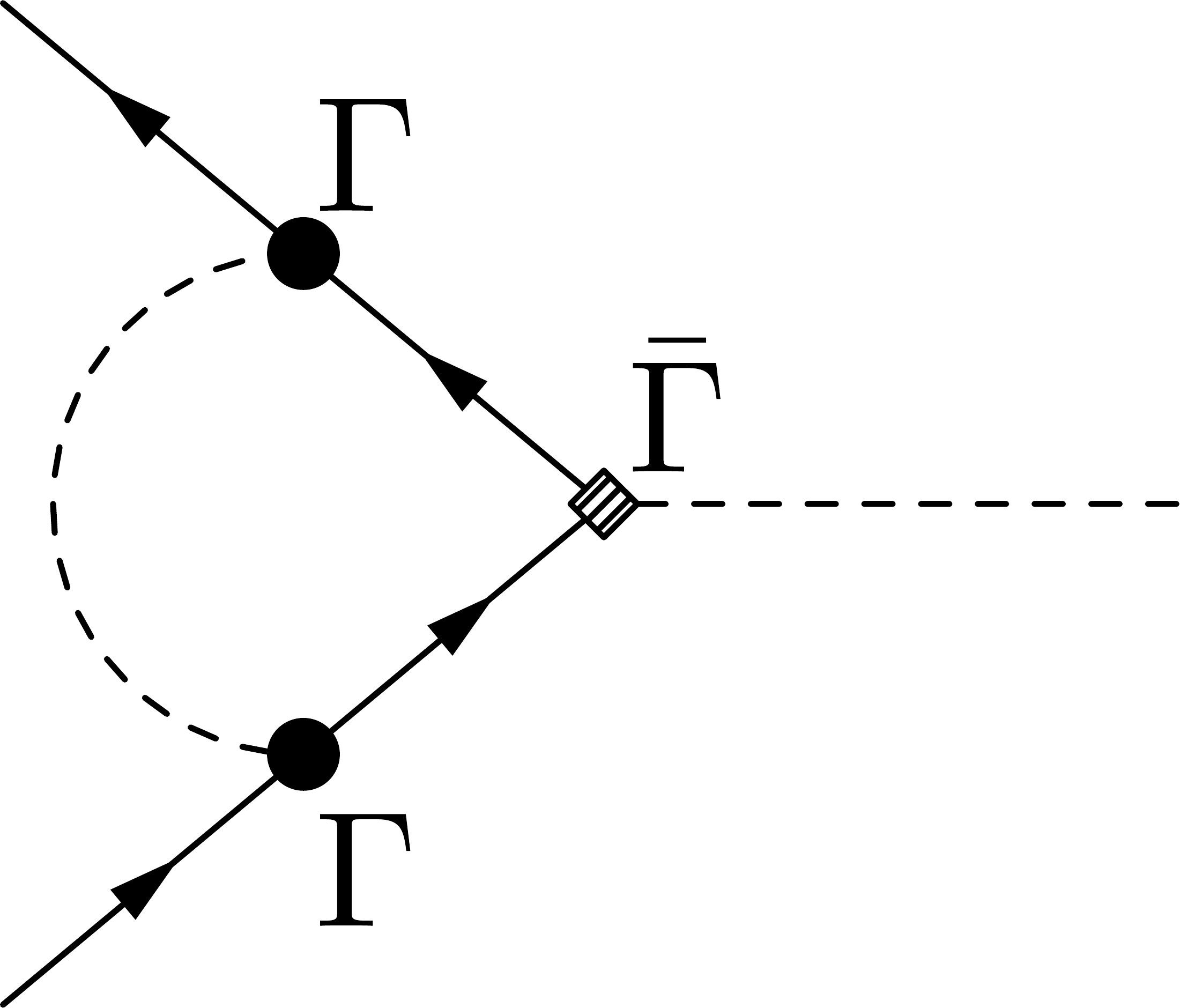}
    \caption{}
    \label{fig:VERTEX_CORR_VAR_COMBI1}
  \end{subfigure}
  \;
  \begin{subfigure}[b]{0.15\textwidth}
  \centering
    \includegraphics[width=0.9\textwidth]{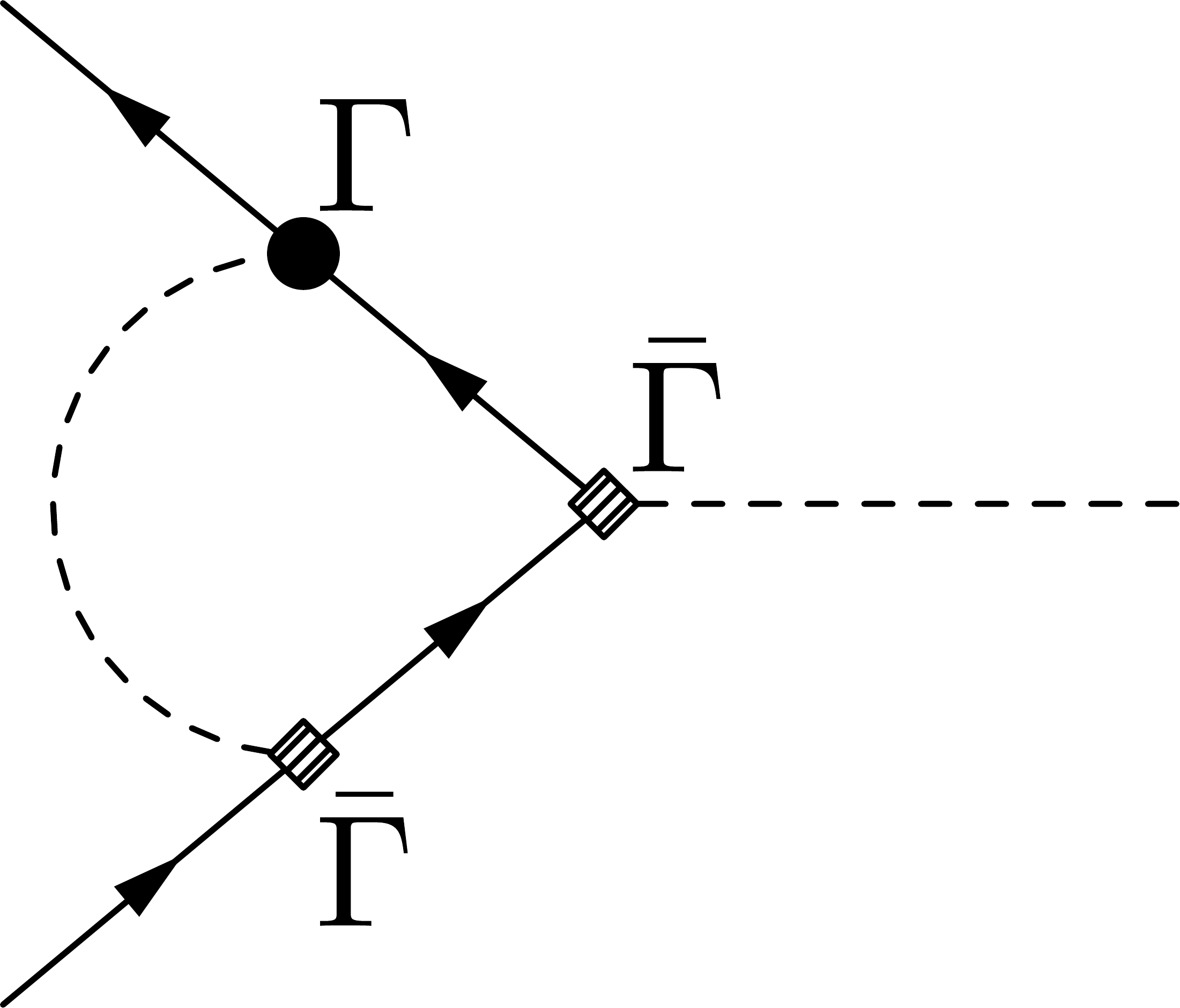}
    \caption{}
    \label{fig:VERTEX_CORR_VAR_COMBI2}
  \end{subfigure}
  \caption{Vertex corrections to disorder couplings $\Gamma, \bar{\Gamma}$.}
  \label{fig:VERTICES}
\end{figure}

We now analyze the flow of the parameters by means of a one-loop expansion in both disorder couplings (note that within our scheme we treat $\lambda$ non-perturbatively). Contributing diagrams are listed in Figs.(\ref{fig:SELFS})-(\ref{fig:VERTICES}), where dotted vertices denote factors $\sim \Gamma$ deriving from Eq.~\eqref{eq:sigma0_vertex_action} and squared vertices indicate factors $\sim \bar{\Gamma}$ coming from Eq.~\eqref{eq:chiral_vertex_action}. We calculate these following a momentum-shell scheme in which we integrate out the Fourier modes within the shell $e^{-l} \Lambda < k _\perp < \Lambda$ and $-\infty <k_\parallel<\infty$ \cite{Roy2016}. Here we restrict ourselves to leading order expressions.

The one-loop contributions to the self-energy are shown in Fig.(\ref{fig:SELFS}). These diagrams are related by straightforward manipulation of external vertices. In particular, Fig.(\ref{fig:SELF}) yields
\begin{align}
\text{Fig.(\ref{fig:SELF})} &= \Gamma^2 \int_{{\bf k}}' \; \; G_{0;\chi}^{t,\lambda} (i \omega, {\bf k}) \\
				&\simeq - i \omega \, \gamma^2  \left( \frac{1 + t \lambda}{1 - t^2} \right) (\sigma_0 - t \, \chi \, {\bf d} \cdot {\pmb \sigma} ) \, l, \nonumber
\end{align}
where we have introduced dimensionless couplings $\gamma^2 = \Gamma^2 \Lambda / (4 \pi v^2 (1 - t^2)^{1/2})$ and $\bar{\gamma}^2 = \bar{\Gamma}^2 \Lambda /( 4 \pi v^2 (1 - t^2)^{1/2})$. Since $( - \chi {\bf d} \cdot {\pmb \sigma})^2 = \sigma_0$ the diagram in Fig.(\ref{fig:SELF_VAR}) differs only by a change in coupling constant,
\begin{align}
\text{Fig.(\ref{fig:SELF_VAR})}	&= \left( \frac{\bar{\Gamma}}{\Gamma} \right)^2 ( - \chi {\bf d} \cdot {\pmb \sigma}) \; \Sigma_a^{t,\lambda} \; ( - \chi {\bf d} \cdot {\pmb \sigma})  \\
				&\simeq - i \omega \, \bar{\gamma}^2  \left( \frac{1 + t \lambda}{1 - t^2} \right) (\sigma_0 - t \, \chi \, {\bf d} \cdot {\pmb \sigma} ) \, l. \nonumber 
\end{align}
The diagram in Fig.(\ref{fig:SELF_COMBI}) has a modified matrix structure,
\begin{align}
\text{Fig.(\ref{fig:SELF_COMBI})}		&= 2 \left( \frac{\bar{\Gamma}}{\Gamma} \right) ( - \chi {\bf d} \cdot {\pmb \sigma}) \;\Sigma_b^{t,\lambda} \\
				&\simeq -2 i \omega \, \gamma \, \bar{\gamma} \left( \frac{1 + t \lambda}{1 - t^2} \right) ( t \sigma_0 - \chi \, {\bf d} \cdot {\pmb \sigma} ) \, l, \nonumber
\end{align}
where we have also taken into account a symmetry factor of two that stems from the freedom to interchange the order of the vertex types. The dressed fermion propagator is then determined from the Dyson equation,
\begin{align}
& G_\chi^{t,\lambda} (i \omega, {\bf k})^{-1} = G_{0;\chi}^{t,\lambda} (i \omega, {\bf k})^{-1} - \Sigma_\chi^{t,\lambda} (i \omega, {\bf k}) \\
&= \left[ i \omega \left( 1 + \frac{1+t \lambda}{1-t^2} \left( (\gamma^2 + \bar{\gamma}^2) + 2t \gamma \bar{\gamma} \right) l  \right) + v t \, {\bf d} \cdot {\bf k} \right] \sigma_0  \nonumber \\
&\; +\chi \, {\pmb \sigma} \cdot\left[ v {\bf k} - i \omega \lambda {\bf d} \left( 1 + \frac{1+t \lambda}{1-t^2} \left( t(\gamma^2 + \bar{\gamma}^2) + 2 \gamma \bar{\gamma} \right) \frac{l}{\lambda} \right) \right]. \nonumber
\end{align}

We similarly calculate the renormalization of the disorder couplings. At one loop order there are two contributing groups of diagrams, each of which has three representatives that are related by manipulation of external vertices. For the first group, Figs.(\ref{fig:VERTEX_CORR})-(\ref{fig:VERTEX_CORR_COMBI2}), we have
\begin{align}
\text{Fig.(\ref{fig:VERTEX_CORR})}	&= \Gamma^3 \int_{\bf k}' \; G_{0;\chi}^{t,\lambda} (0, {\bf k})^2 \\
						&\simeq \Gamma \, \gamma^2 \left( \frac{1}{1 - t^2} \right) (\sigma_0 - t \, \chi \, {\bf d} \cdot {\pmb \sigma} ) \,l. \nonumber \\
\text{Fig.(\ref{fig:VERTEX_CORR_COMBI1})} 	&= \left( \frac{\bar{\Gamma}}{\Gamma} \right)^2 \, ( - \chi {\bf d} \cdot {\pmb \sigma}) \, \left[\text{Fig.(\ref{fig:VERTEX_CORR})} \right] \, ( - \chi {\bf d} \cdot {\pmb \sigma}) \\
								&\simeq \Gamma \, \bar{\gamma}^2 \left( \frac{1}{1 - t^2} \right) (\sigma_0 - t \, \chi \, {\bf d} \cdot {\pmb \sigma} ) \,l, \nonumber \\
\text{Fig.(\ref{fig:VERTEX_CORR_COMBI2})} 	&= 2 \left( \frac{\bar{\Gamma}}{\Gamma} \right) \, ( - \chi {\bf d} \cdot {\pmb \sigma}) \, \left[\text{Fig.(\ref{fig:VERTEX_CORR})} \right] \\
								&\simeq 2 \Gamma \, \gamma \, \bar{\gamma} \left( \frac{1}{1 - t^2} \right) (t \, \sigma_0 - \chi {\bf d} \cdot {\pmb \sigma} ) \,l, \nonumber
\end{align}
where we have again accounted for a symmetry factor of two for the last diagram. The diagrams of the second group, (\ref{fig:VERTEX_CORR_VAR})-(\ref{fig:VERTEX_CORR_VAR_COMBI2}), are given by
\begin{align}
\text{Fig.(\ref{fig:VERTEX_CORR_VAR})}	&= \bar{\Gamma}^3 \int_{\bf k}' \; ( - \chi {\bf d} \cdot {\pmb \sigma}) \left( G_{0;\chi}^{t,\lambda} (0, {\bf k}) ( - \chi {\bf d} \cdot {\pmb \sigma}) \right)^2 \nonumber \\
							&\simeq \bar{\Gamma} \, \bar{\gamma}^2 \left( \frac{t}{1 - t^2} \right) (\sigma_0 - t \, \chi \, {\bf d} \cdot {\pmb \sigma} ) \,l. \\
\text{Fig.(\ref{fig:VERTEX_CORR_VAR_COMBI1})} 	&= \left( \frac{\Gamma}{\bar{\Gamma}} \right)^2 \, ( - \chi {\bf d} \cdot {\pmb \sigma}) \, \left[\text{Fig.(\ref{fig:VERTEX_CORR_VAR})} \right] \, ( - \chi {\bf d} \cdot {\pmb \sigma}) \nonumber \\
								&\simeq \bar{\Gamma} \, \gamma^2 \left( \frac{t}{1 - t^2} \right) (\sigma_0 - t \, \chi \, {\bf d} \cdot {\pmb \sigma} ) \,l,  \\
\text{Fig.(\ref{fig:VERTEX_CORR_VAR_COMBI2})} 	&= 2 \left( \frac{\Gamma}{\bar{\Gamma}} \right) \, ( - \chi {\bf d} \cdot {\pmb \sigma}) \, \left[\text{Fig.(\ref{fig:VERTEX_CORR_VAR})} \right] \nonumber \\
								&\simeq 2 \bar{\Gamma} \, \gamma \, \bar{\gamma} \left( \frac{t}{1 - t^2} \right) (t \, \sigma_0 - \chi {\bf d} \cdot {\pmb \sigma} ) \,l. 
\end{align}

With point like disorder, Eq.(\ref{eq:disorder_corr}), the self-energy is independent of external momentum, meaning there is no direct renormalization of the Fermi velocity; it is just given by the inverse fermion field rescaling~\cite{Goswami2011,Roy2014}. The flow of the dimensionless disorder couplings is due to vertex corrections only. Including the bare scaling dimensions of the parameters we find leading-order $\beta$-functions
\begin{align}
\beta_v			&= -\frac{1+t \lambda}{1-t^2} \left( (\gamma^2 + \bar{\gamma}^2) + 2t \gamma \bar{\gamma}  \right) v, \label{eq:Velocity_beta_function} \\
\beta_\gamma		&= - \frac{ \gamma}{2} + \frac{\gamma + t \bar{\gamma}}{ 1 - t^2} \left( (\gamma^2 + \bar{\gamma}^2) + 2t \gamma \bar{\gamma} \right), \label{eq:Disorder_beta_function} \\
\beta_{\bar{\gamma}}	&= - \frac{ \bar{\gamma}}{2} + \frac{ \gamma + t \bar{\gamma}}{ 1 - t^2} \left( t(\gamma^2 + \bar{\gamma}^2) + 2 \gamma \bar{\gamma} \right), \label{eq:Anomalous_disorder_beta_function} \\
\beta_\lambda		&= \frac{1+t \lambda}{1-t^2} \left( (t - \lambda) (\gamma^2 + \bar{\gamma}^2) + 2 (1 - t \lambda) \gamma \bar{\gamma} \right), \label{eq:Lambda_beta_function}
\end{align}
where $\beta_\alpha=\mathrm{d} \alpha / \mathrm{d} l$ with $\alpha=v,\gamma,\bar{\gamma},\lambda$. 

\begin{figure}[!htb]
\centering
  \begin{subfigure}[t]{0.45\textwidth}
	\centering
	\includegraphics[width=0.975\textwidth]{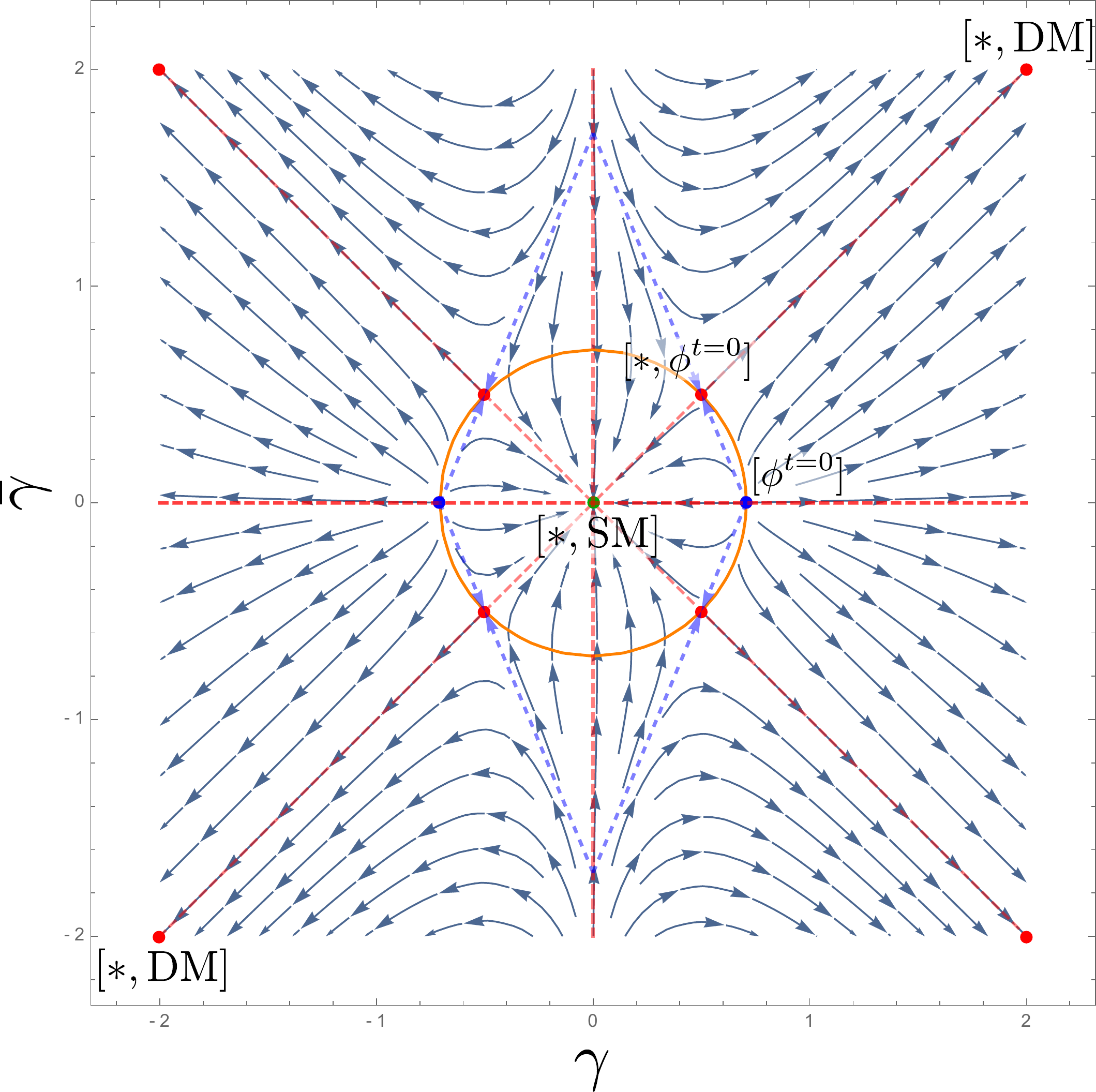}
	\caption{$t=0$.}
	\label{fig:Flow_disorder_untilted_SCBA}
   \end{subfigure} 
  \begin{subfigure}[t]{0.45\textwidth}
	\centering
	\includegraphics[width=0.975\textwidth]{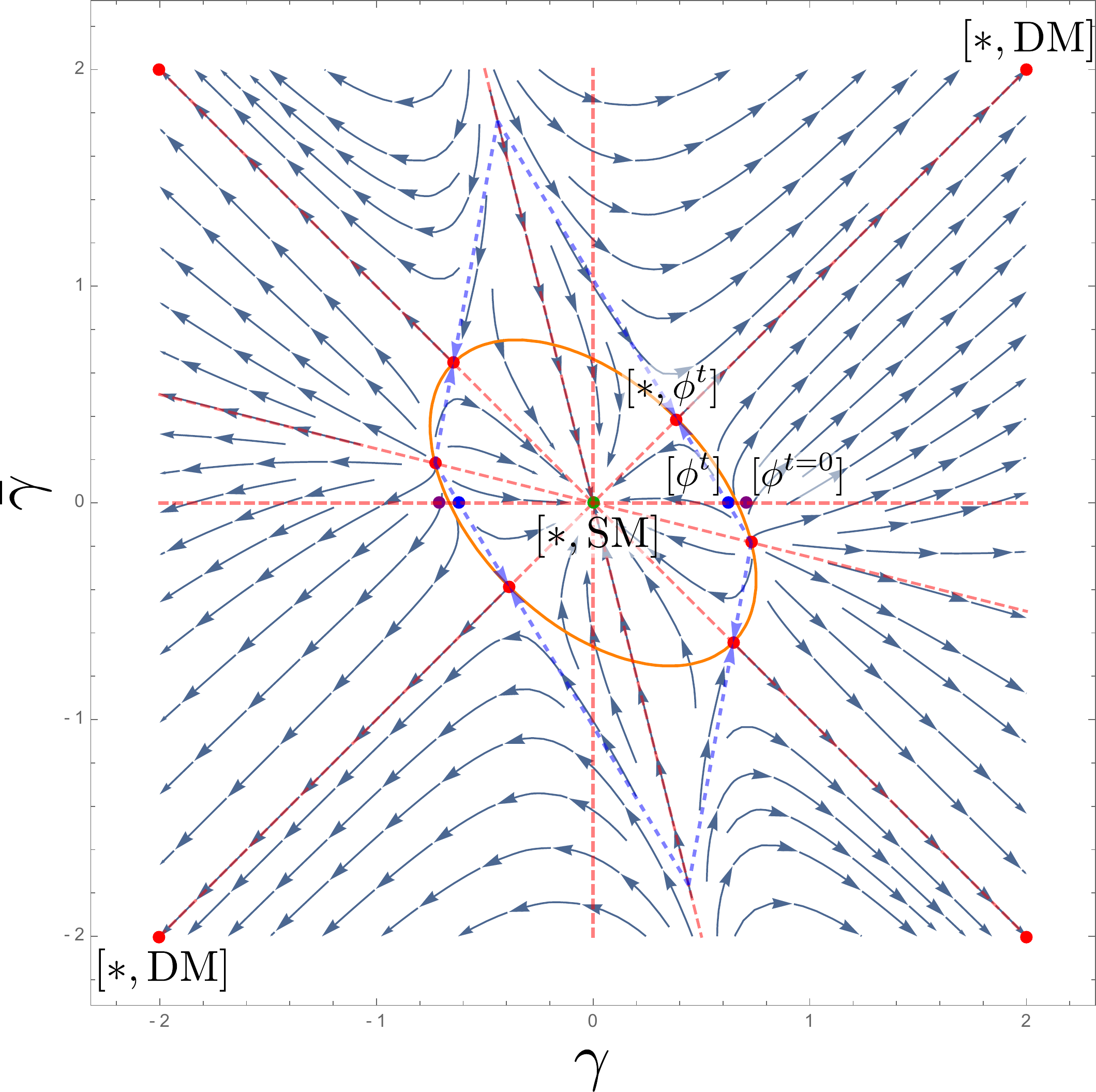}
	\caption{$t=0.25$.}
	\label{fig:Flow_disorder_pos_tilt_SCBA}
   \end{subfigure} 
  \caption{Flow for disorder couplings $\gamma, \bar{\gamma}$ and \textcolor{YellowOrange}{SCBA }elliptic critcal disorder line (see appendix \ref{appsec:scba})  in (a) the untilted case, $t=0$ and (b) the tilted case, $t=0.25$. \textcolor{red}{Red} dots: disorder fixed points. \textcolor{blue}{Blue} dots: phase transition points. \textcolor{DarkOrchid}{Purple} dot: phase transition fixed point $(\gamma, \bar{\gamma})_{[\phi^{t=0}]} = (1/ \sqrt{2},0)$. }
  \label{fig:Disorder_flow}
\end{figure}

\subsection{Discussion of the RG equations}

Under rescaling the model's parameters will flow away from their initial values, which are indicated here by subscript $0$. Note that the starting values of the generated parameters are $\lambda_0=\bar{\gamma}_0=0$. There are multiple parameter combinations for which $\beta$-functions (\ref{eq:Velocity_beta_function})-(\ref{eq:Lambda_beta_function}) vanish simultaneously and the flow terminates. These fixed points are indicated by subscript asterisk $*$. A first observation is that the tilt parameter $t$ is unaffected by the flow meaning there is no renormalization of the tilt within our renormalization scheme. As a consequence, the equations (\ref{eq:Disorder_beta_function})-(\ref{eq:Anomalous_disorder_beta_function}) for the dimensionless couplings $\gamma$ and  $\bar{\gamma}$ form a closed subset, producing a purely two-dimensional flow that is shown in Fig.(\ref{fig:Disorder_flow}). 

\begin{figure}[h!]
	\centering
	\includegraphics[width=0.42\textwidth]{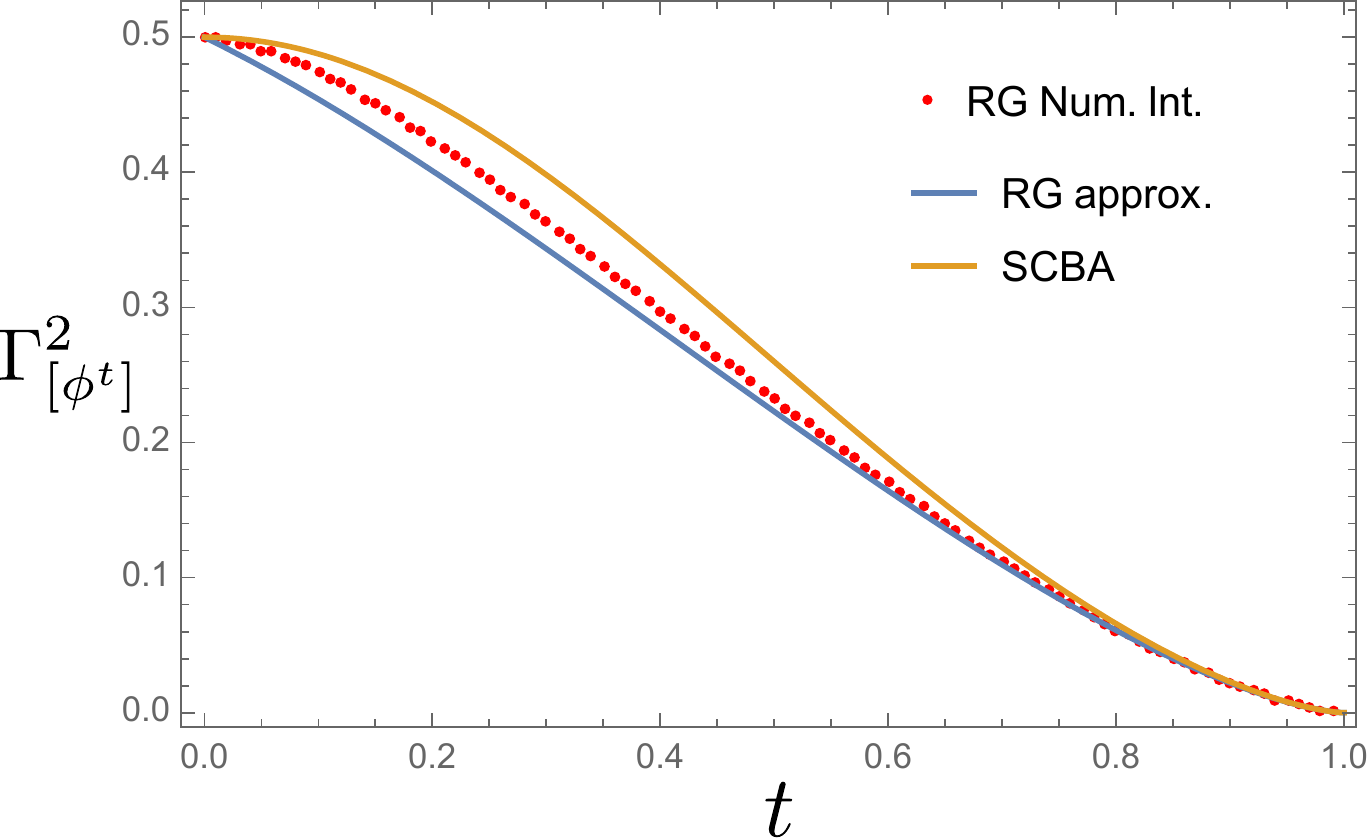}
	\caption{Dependence of the phase transition disorder strength $\Gamma_{[\phi]}$ on the ordinary tilt $t$. The \textcolor{YellowOrange}{SCBA} (see appendix (\ref{appsec:scba})) consistently overestimates the disorder required for the WSM-DM transition, whereas the boundary of the approximate \textcolor{blue}{weak-disorder parallelogram} lead to a lower estimate.} 
	\label{fig:Tilt_v_crit_disorder}
\end{figure}

{\it Disorder flow:} In general, there are three distinct fixed points that can be reached when initiating the flow at $\bar{\gamma}_0 =0$. First of all, there is the attractive weak-coupling fixed point $(\gamma, \bar{\gamma})_{[*,\text{SM}]} = (0,0)$ which is associated with semimetallic behavior, called SM. Completely oppositely sit the attractive symmetry-related strong-disorder fixed points $(\gamma, \bar{\gamma})_{[*,\text{DM}]} = \pm(\infty,\infty)$, where the disorder strengths diverge. More interestingly, there are also the symmetry-related intermediate coupling fixed points
\begin{align} \label{eq:intermediate_fixed_point}
(\gamma, \bar{\gamma})_{[*,\phi^t]}= \pm \frac{1}{2} \left( \sqrt{\frac{1-t}{1+t}}, \sqrt{\frac{1-t}{1+t}} \right),
\end{align}
 that are attractive in two out of four directions and repulsive in the others. These fixed points will move towards the origin on the line $\gamma = \bar{\gamma}$ as $t$ is increased, eventually merging with the weak-disorder fixed point $[*,\text{SM}]$ for $t=1$. The flow of the conventional untilted case $t=0$ is presented in Fig.(\ref{fig:Flow_disorder_untilted_SCBA}). The disorder $\beta$-functions reduce to $\beta_{\gamma} = \gamma ( \gamma^2 + \bar{\gamma}^2) - \gamma/2$ and $\beta_{\bar{\gamma}} =  2 \bar{\gamma} \gamma^2 - \bar{\gamma}/2$, meaning $\bar{\gamma}$ is never generated and the flow remains effectively one-dimensional if the flow is initiated from $\bar{\gamma}_0$. This means that the intermediate fixed point $[*,\phi^t]$ is never accessed in the untilted model. It however allows for an unstable fixed point at $(\gamma, \bar{\gamma})_{[\phi^{t=0}]} = (0, 1 / \sqrt{2})$ at which a transition from the weak-disorder into the strong-disorder regime occurs \cite{Goswami2011,Roy2014,Altland2016}. For finite tilts $t > 0$, the flow in Fig.(\ref{fig:Flow_disorder_pos_tilt_SCBA}) is fundamentally different. It holds on the horizontal axis $\bar{\gamma}=0$ that $\beta_{\bar{\gamma}} = t \gamma^3/(1 - t^2)$. Some $\bar{\gamma}$ is spontaneously generated in any disordered tilted Weyl system, which destabalizes the untilted fixed point. In return, the intermediate fixed point $[*,\phi]$ in Eq.(\ref{eq:intermediate_fixed_point}) is now accesible to flow. It is reached only exactly from the phase transition point $[\phi^t]$ at $(\gamma, \bar{\gamma})_{[\phi^t]} = (\gamma_{[\phi^t]},0)$. We stress that $[\phi^t]$ is not an fixed point of the theory for finite tilts, but merely the point where the flow changes direction for $\bar{\gamma}=0$.  To the left of $[\phi^t]$ on the horizontal axis the flow is towards weak disorder, whereas flow emanating form the right is directed towards strong disorder. Comparing Fig.(\ref{fig:Flow_disorder_untilted_SCBA}) and Fig.(\ref{fig:Flow_disorder_pos_tilt_SCBA}) supports the observation \cite{Trescher2017} that including a finite tilt reduces the phase transition disorder strength, $\gamma_{[\phi^t]} \leq \gamma_{[\phi^{t=0}]}$ . We substantiate this conclusion in Fig.(\ref{fig:Tilt_v_crit_disorder}), which directly graphs the dependency of the original phase transition disorder strength $\Gamma_{[\phi^t]}$ on $t$. Note that Fig.(\ref{fig:Tilt_v_crit_disorder}) also contains a line which is obtained from a fully analytical solution of the self-consistent Born approximation (SCBA) which is presented in Appendix~\ref{appsec:scba} and is consistent with an earlier analysis~\cite{Trescher2017}.

\vspace{12pt}

Concentrating on flow emanating from $\bar{\gamma}_0=0$, the asymptotic relation of the disorder couplings is surmised by $\bar{\gamma}_* = \gamma_*$ regardless of the side of the transition we study. We can then investigate the terminal behavior of the other parameters by restricting to the line $\bar{\gamma} = \gamma$. 

\begin{figure}[!h]
\centering
  \begin{subfigure}[t]{0.38\textwidth}
	\centering
	\includegraphics[width=0.975\textwidth]{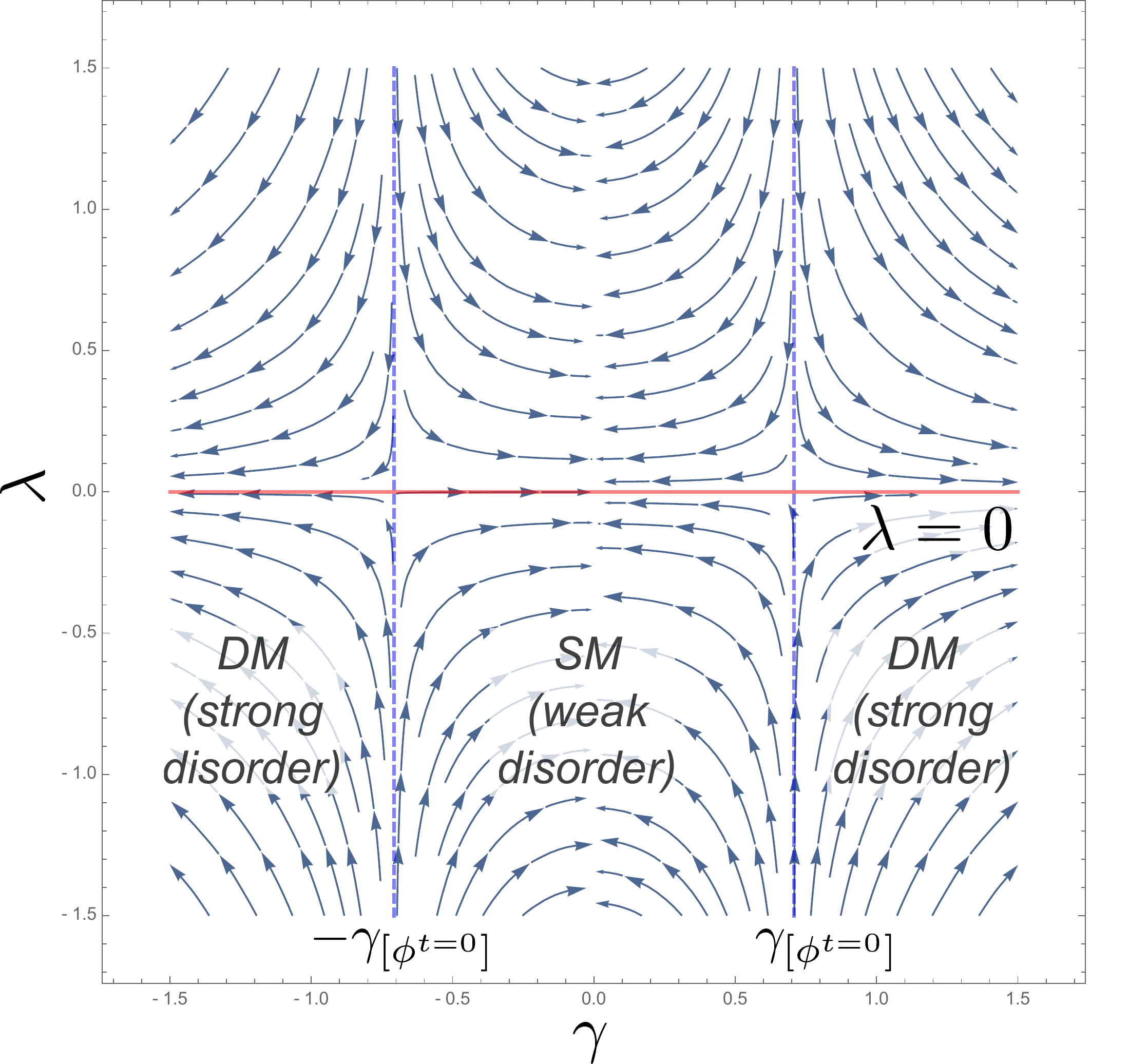}
	\caption{$t=\bar{\gamma}=0$.}
	\label{fig:Untilted_lambda_flow}
   \end{subfigure} 
  \begin{subfigure}[t]{0.38\textwidth}
	\centering
	\includegraphics[width=0.975\textwidth]{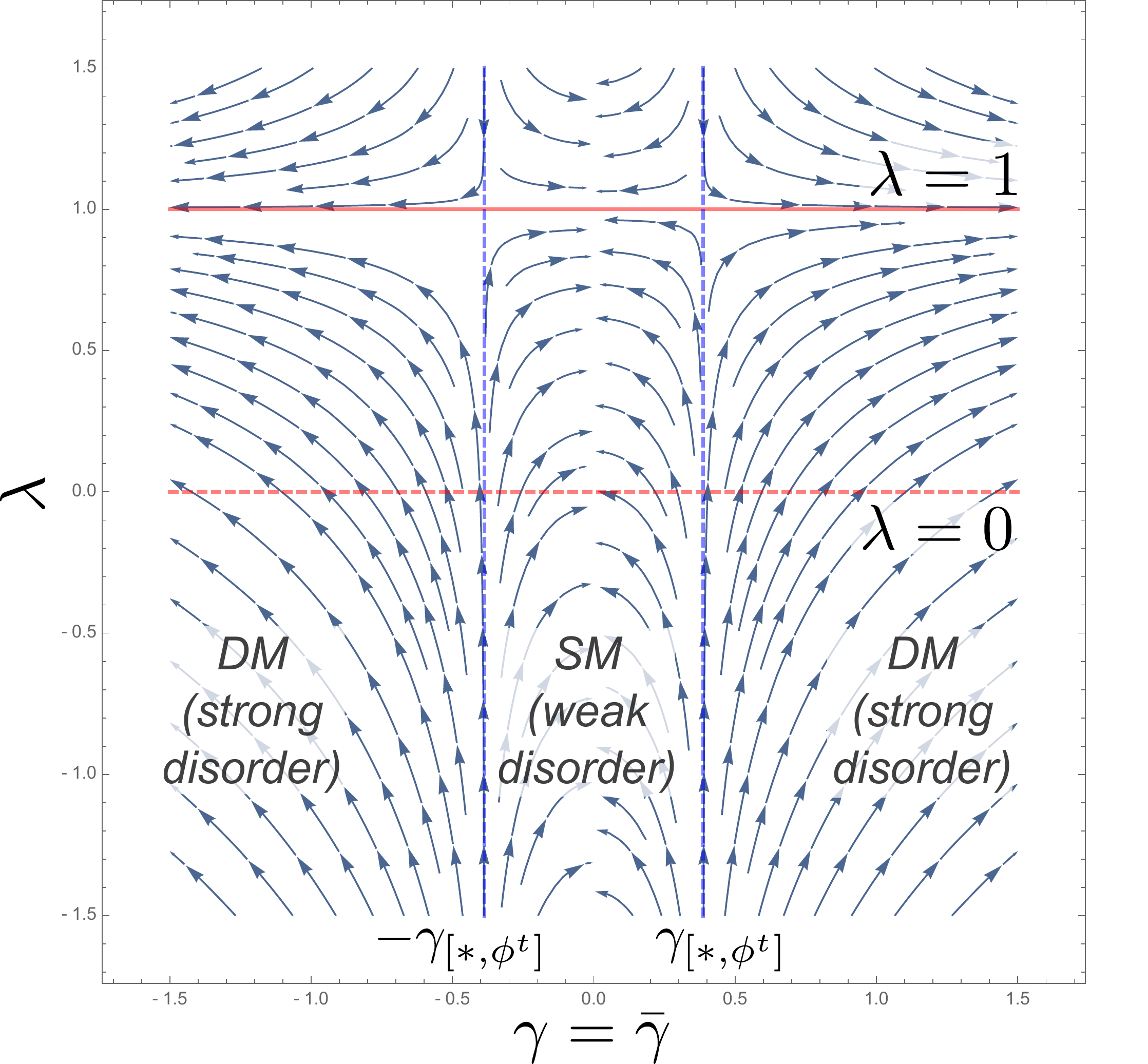}
	\caption{$t=0.25$, $\gamma=\bar{\gamma}$.}
	\label{fig:Pos_tilted_lambda_flow}
   \end{subfigure} 
  \caption{Streamplot of anomalous tilting versus disorder for the appropriate terminal relations of $\gamma$ and $\bar{\gamma}$. The flow predicts that $\lambda$ will find a terminal value $0 \leq \lambda_* \leq 1$. }
  \label{fig:Lambda_flow}
\end{figure}

\begin{figure}[h!]
	\centering
	\includegraphics[width=0.375\textwidth]{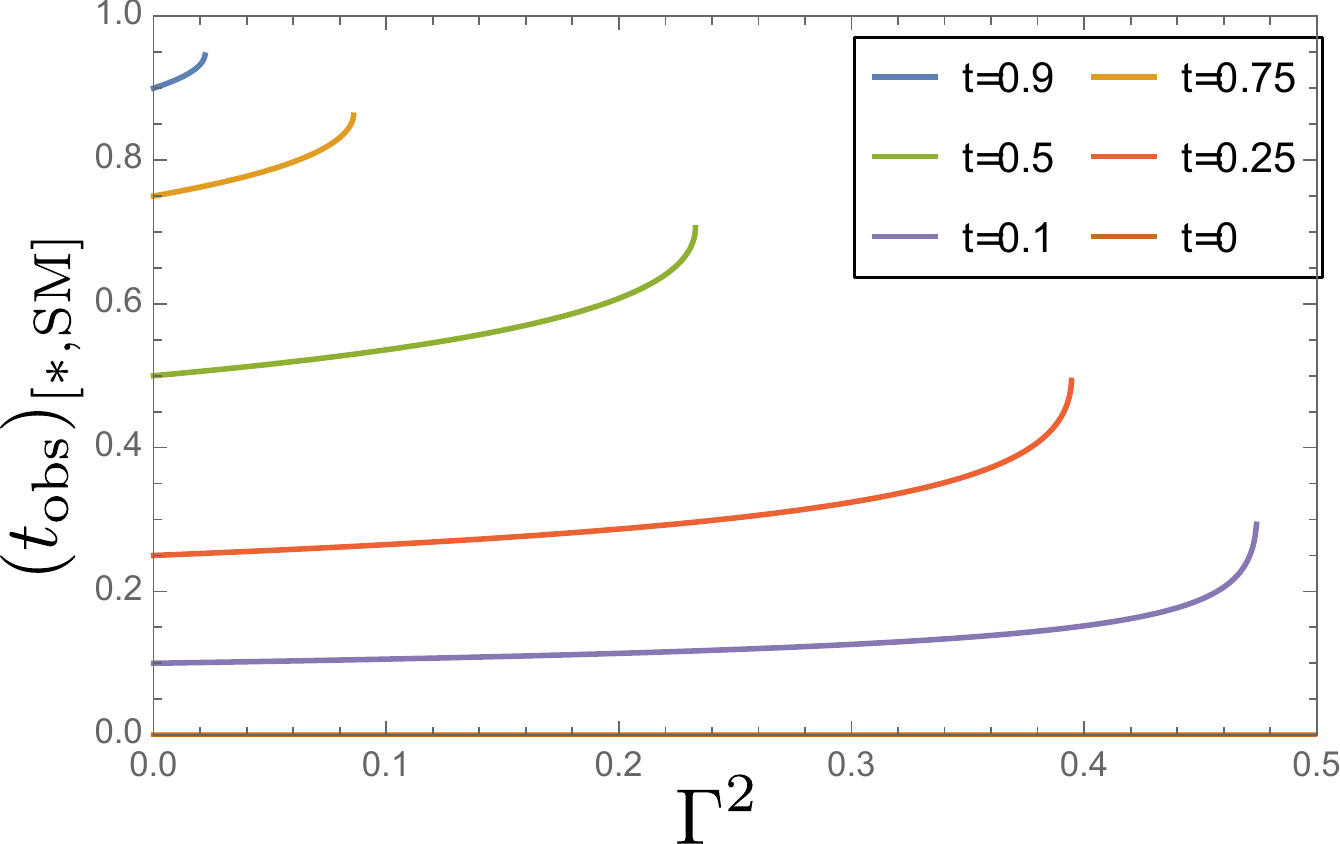}
	\caption{Terminal observable tilt at the weak disorder fixed point $(t_\text{obs})_{[*,\text{SM}]}$ reached as a function of disorder strength $\Gamma$ for selected value of original tilt $t$.} 
	\label{fig:Terminal_obs_tilt_function}
\end{figure}

{\it Anomalous tilt flow:} Insertion of the terminal relation $\bar{\gamma} = \gamma$ into Eq.(\ref{eq:Lambda_beta_function}) yields the simplified $\lambda$-flow equation $\beta_\lambda = 2 \gamma^2 (1  - \lambda) (1+t \lambda) / (1 - t)$ valid on this line. Depending on the original tilt there are now two distinct scenarios that can be realized in a disordered system, also summarized in Fig.(\ref{fig:Lambda_flow}):
\begin{itemize}
\item[(1).] for $t=0$ we have that $\beta_\lambda = - 2 \lambda \gamma^2$, so that initial value $\lambda_0=0$ means that $\lambda$ will never be generated.
\item[(2).]  for $t>0$ the anomalous tilt increases from $\lambda_0=0$ until reaching a fixed point with  $0 < \lambda_{[*,\text{SM}]} < 1$ inside the weak-disorder region and $\lambda_{[*,\text{DM}]} = 1$ outside of it.
\end{itemize}
In general the final anomalous tilt $\lambda_*$ after following the flow is a monotonically increasing function both of tilt $t$ and disorder strength $\gamma_0$. $\lambda$ will spontaneously acquire a finite value for any tilt $0 < t < 1$, even within the weak-disorder region. By Eq.(\ref{eq:obs_tilt}) it follows that disorder {\it indirectly} increases the observable tilt in the system that is reached after concluding the disorder flow, although a final value $(t_\text{obs})_*=1$ is never reached. This restricted growing behavior is depicted in Fig.(\ref{fig:Terminal_obs_tilt_function}). Earlier results \cite{Trescher2017} based on a more limited SCBA are qualitatively the same, but fail to pick up on the noticable uptick in $(t_\text{obs})_{[*,\text{SM}]}$ close to phase transition disorder strength. On the level of observables like the renormalized DoS, the transition from type-I to type-II Weyl cone is concealed by the disorder-induced WSM-DM transition. 

{\it Velocity flow:} Similarly, entering $\bar{\gamma} = \gamma$ into Eq.(\ref{eq:Velocity_beta_function}) yields the velocity flow equation $\beta_v = - 2 v \gamma^ 2  (1+t \lambda) / (1 - t)$ valid on this line. Note that, since we have verified that $0 <\lambda_* \leq 1$, it holds that $\beta_v$ cannot be positive. Thus, $v$ decreases until reaching a fixed point with a lower but finite Fermi velocity $0 < v_{[*,\text{SM}]} < v_0$ inside the weak-disorder region and $v_{[*,\text{DM}]} =0$ outside of it.

\vspace{12pt} 

Note that the  complete set of $\beta$-functions Eqs.(\ref{eq:Velocity_beta_function})-(\ref{eq:Lambda_beta_function}) can be numerically integrated directly to study the behavior of the model under flow for a given set of initial parameter values. One might check that doing so confirms the analytical considerations above and substantiates the derived conclusions.

\subsection{Interpretation and Discussion} 

The $\beta$-functions Eqs.(\ref{eq:Velocity_beta_function})-(\ref{eq:Lambda_beta_function}) that result from RG treatment of disorder in the (anomalously) tilted Weyl cone produce a flow that can be divided into two qualitatively different regions separated by a quantum phase transition. In the weak-disorder region the flow of $\gamma$ and $\bar{\gamma}$ is directed towards the origin. Under sufficient rescalings to smaller energy scales the system reaches the clean limit, albeit with lower but finite Fermi velocity $v_{[*,\text{SM}]}$ and at the price of acquiring a finite anomalous contribution $0 < \lambda_{[*,\text{SM}]} < 1$ to the tilt. It will thus preserve its semimetallic properties, such as a density of states that continues to vanish at the Weyl point and grow quadratically away from it along the lines of Eq.(\ref{eq:DoS_obs}). 

This behavior changes as the strength of the disorder in the model nears the phase transition point $[\phi^t]$. There, disorder flow is directed towards the intermediate fixed point $[*,\phi^t]$ of Eq.(\ref{eq:intermediate_fixed_point}), at which the velocity and anomalous tilt have $\beta$-functions 
\begin{align*}
\beta_v |_{\,_{[*,\phi^t]}} = - \frac{v(1 + t \lambda)}{2(1+t)}, \qquad  \beta_\lambda |_{\,_{[*,\phi^t]}} =  \frac{(1-\lambda)(1 + t \lambda)}{2(1 + t)}
\end{align*}
that show that their flow terminates for $v_{[*,\phi^t]} = 0$ and $\lambda_{[*,\phi^t]} = 1$. However, as mentioned before this does not imply that the system goes metallic. The residual scaling of $v$ and $\lambda$ at the fixed point $\phi$ is borne out also in the DoS, which has the flow equation $\beta_{\rho^{t,\lambda}} |_{[*,\phi^t]} = \rho^{t,\lambda}$ by virtue of its flowing parameters. Integration of this equation at cutoff scales that have become comparable to the chemical potential show that the disordered DoS will develop an additional inverse power of $\omega$ at criticality. Although the system remains semimetallic, its DoS is enhanced to scale linearly,
\begin{align} \label{eq:dis_damp_DoS}
\rho^{t,\lambda}  |_{\,_{[*,\phi^t]}} (\omega) \sim \frac{\omega} {(1- t^2)^2 v_0^3}.
\end{align}
It is worth noting that this yields critical exponents that are indistinguishable from those of the disordered untilted cone \cite{Ominato2014,Roy2014,Roy2016}, but has substantially different prefactors.

On the other side of the transition sits a phase with entirely different characteristics. In the strong-disorder region both $\gamma$ and $\bar{\gamma}$ undergo divergent flow signaling a breakdown of perturbation theory. Such problems can be circumvented by other methods such as the self-consistent Born approximation (SCBA) that we apply in appendix (\ref{appsec:scba}): as an ansatz one supplements the Green function with some imaginary self-energy, which introduces a new length scale into the model with which to compare the cutoff. One then studies the conditions for which this self-energy is self-consistently allowed to take on non-zero values. Because finite imaginary self-energies must necessarily give a non-vanishing contribution to the density of states, i.e. $\rho (0) > 0$, this by definition can only take place in the strong-disorder region. The finite density of states observed even at the energy of the Weyl point shows that the strong-disorder phase is metallic in nature. 

For generic three-dimensional models under disorder effects it is commonly accepted that there is a mobility edge separating localized states from extended states. The associated phase transition connects the diffusive metal with states above the mobility edge to the Anderson insulator on the other side \cite{Anderson1958,Abrahams1979}. However, the model currently studied here contains only a single isolated Weyl cone and thus precludes the coherent backscattering thought to be a necessary precondition for Anderson localization \cite{Bardarson2007}. Such processes can be included by connecting two copies of the single Weyl fermion model with opposite chirality via off-diagonal intercone-scattering. The resulting Dirac-type model, with a four-dimensional matrix structure that allows for the spontaneous generation of yet more unexpected terms, will be the subject of a forthcoming paper. Another outstanding question relates to  the stability of the disorder-induced WSM-DM transition to interactions. Coulomb-type interactions tend to renormalize the tilt towards zero and cause the model to flow towards the untilted limit at low energies \cite{Detassis2017}, thus competing with the disorder-mediated effects studied here. 

\section{Conclusion} \label{sec:conc}

Within this paper we studied disorder in a model of a single tilted Weyl cone of type-I starting from a weak coupling perspective. We find a complicated multiparameter phase diagram which however still harbors a quantum critical point between a semimetallic and a diffusive metallic phase. We find that, like in the untilted case, as the system approaches critical disorder strength, the scaling of the density of states is enhanced to grow linearly with energy. We furthermore show within a renormalizaiton group approach that finite tilt quickens the disorder-induced semimetal-metal transition, see Fig.(\ref{fig:Tilt_v_crit_disorder}), and that disorder renormalizes the observable tilt to increase, see. Fig.(\ref{fig:Terminal_obs_tilt_function}), consistent with earlier work on this subject~\cite{Trescher2017}. 
Future directions of research are to verify our findings numerically using the kernel polynomial method and also investigate the robustness under the addition of another cone. 
It is also interesting to contrast the increased tilt found in the disordered system from a recent finding in interacting Weyl systems where Coulomb interaction was shown to decrease tilt~\cite{Detassis2017}. Therefore, one should expect an interesting interplay between the two.  

\vspace{12pt}
{\it{Acknowledgment:}}
T.S.S. thanks E. van der Wurff and M. Trescher for useful conversations. L. F. acknowledges discussions with F. Detassis and S.Grubinskas on related works. The authors furthermore acknowledge discussions with P.W. Brouwer, V. Juri\v{c}i\'{c}, B. Roy, and B. Sbierski. This work is part of the D-ITP consortium, a program of the Netherlands Organisation for Scientific Research (NWO) that is funded by the Dutch Ministry of Education, Culture and Science (OCW). 

\newpage

\renewcommand{\thesection}{\Alph{section}}
\setcounter{equation}{0}  
\setcounter{section}{0}
\setcounter{subsection}{0}

\numberwithin{equation}{section}

\newpage
\section{Self-Consistent Born Approximation (SCBA)} \label{appsec:scba}

Besides the RG analysis of the main body we also studied the disorder-induced semimetal-diffusive metal (SM-DM) transition in tilted Weyl cones by means of a self-consistent Born approximation (SCBA). Implementing the ansatz $\Sigma_\chi^\text{SCBA} = i \alpha \, \sigma_0 - i \, \chi \, \beta \, {\bf d} \cdot {\pmb \sigma}$ means that the self-energy acquires a finite imaginary part for $\alpha,\beta > 0$, which is directly related to the emergence of a finite DoS even at the nodal point across the transition \cite{Ominato2014,Altland2016,Trescher2017}. We can then look for such solutions self-consistently and express the conditions for their emergence in terms of the critical disorder couplings. 

\begin{figure}[h]
\centering
\includegraphics[width=0.35\textwidth]{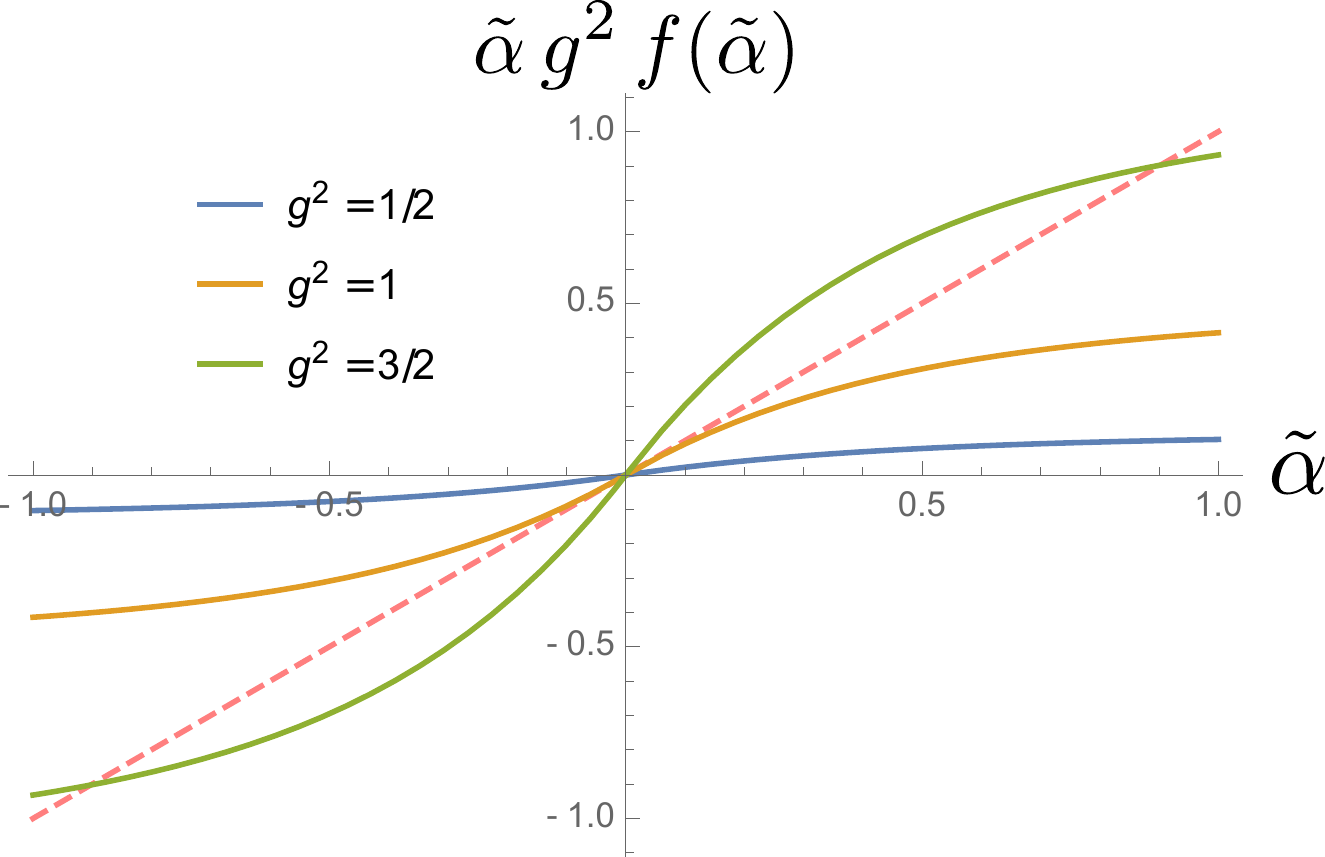}
  \caption{(a)-(c): Comparison of $\tilde{\alpha}$ and $\tilde{\alpha} \, g^2 \, f(\tilde{\alpha} )$ for the indicated composite-couplings $g$.}
  \label{fig: FUNCTION}
\end{figure}

Inspired by the result of the RG equations we calculate the self-energy with a more generic disorder term $ \Upsilon_{\Gamma, \bar{\Gamma}} = (\Gamma \sigma_0 \mp \bar{\Gamma} \, {\bf d} \cdot {\pmb \sigma})$ that is added to the non-interacting model corresponding to Eq.(\ref{eq:ham}). This yields the self-consistency equation
\begin{align}
&\Sigma_\chi^\text{SCBA} = \int'_{\bf k} \;  \Upsilon_{\Gamma, \bar{\Gamma}}  \, \left[ G_{0;\chi}^{t,\lambda} (0, {\bf k})^{-1} - \Sigma_\chi^\text{SCBA}  \right]^{-1} \,  \Upsilon_{\Gamma, \bar{\Gamma}}  \nonumber \\
& \; = i \left( \Big( (\gamma^2 + \bar{\gamma}^2) + 2 t \gamma \bar{\gamma} \Big) \sigma_0 - \chi  {\bf d} \cdot {\pmb \sigma} \Big( t (\gamma^2 + \bar{\gamma}^2) + 2 \gamma \bar{\gamma} \Big) \right) \nonumber  \\
& \qquad \qquad \times (1 - t^2)^{-1} \, (\alpha + t \beta) \; f  (\alpha + t \beta), \nonumber 
\end{align}
where $f$ is a complicated function of $\alpha + t \beta $. Due to the ansatz matrix structure of the self-energy, this is solved only by a constant ratio
\begin{align}
\rho = \frac{\beta}{\alpha} =  \frac{ t (\gamma^2 + \bar{\gamma}^2) + 2 \gamma \bar{\gamma} } { (\gamma^2 + \bar{\gamma}^2) + 2 t \gamma \bar{\gamma}}. 
\end{align}
Defining $\tilde{\alpha}^2 = (1  + t \rho)^2 \alpha^2$, we then find a scalar equation for this remaining unknown:
\begin{align} \label{eq:SCBA_II}
\tilde{\alpha} = \tilde{\alpha} \, g^2  f \left( \tilde{\alpha} \right), 
\end{align}
where $g^2 = (1 + t \rho) /( 1 - t^2) ( (\gamma^2 + \bar{\gamma}^2) + 2 t \gamma \bar{\gamma})$ is a suitably defined composite coupling. We can now determine the qualitative solutions to the self-consistency equation for $\tilde{\alpha}$ by contrasting the LHS and RHS of Eq.(\ref{eq:SCBA_II}), see Fig.(\ref{fig: FUNCTION}). Indeed, we find that for $g^2 \leq 1$ the equation is solved only by trivially setting $\tilde{\alpha}=0$, whereas $g^2 >1$ opens up the possibility of a finite self-energy. Thus, we find an exact critical disorder line in coupling space that is determined by the relation
\begin{align} \label{eq:SCBA_crit_disorder_ellipse}
(1 - t^2) = (1 + t^2) (\gamma^2 + \bar{\gamma}^2) + 4 t \gamma \bar{\gamma} \;.
\end{align}
This equation describes an elliptic critical disorder line that coincides exactly with all the non-trivial disorder fixed points, see Fig.(\ref{fig:Flow_disorder_pos_tilt_SCBA}), as long as we take into account a known factor of two difference between SCBA and renormalisation group results \cite{Trescher2015,Trescher2017}. From the same picture, we can deduce that SCBA analysis slightly overestimates the critical disorder strength $\gamma_c^t$, with the terminal direction of the flow switching only between the blue and yellow point on the positive branch of the $\bar{\gamma}=0$ axis.

We extract the critical disorder strength as a function of tilt $t$ at which the system transitions from semimetal to diffusive metal from the SCBA ellipse defined by Eq.(\ref{eq:SCBA_crit_disorder_ellipse}). Setting $\bar{\gamma}=0$ we find that it is given by
\begin{align}
 2(1+t^2) (\gamma_\phi^t)^2 = (1-t^2),
\end{align}
where again we have accounted for a factor of $2$. This exact relation is to be compared with the results gathered from RG analysis, see Fig.(\ref{fig:Tilt_v_crit_disorder}). Scaling back to original coupling $\Gamma$, this graph reproduces the results from Ref.\cite{Trescher2017} up to normalization. 

\vspace{12pt}

\begin{thebibliography}{40}%
\makeatletter
\providecommand \@ifxundefined [1]{%
 \@ifx{#1\undefined}
}%
\providecommand \@ifnum [1]{%
 \ifnum #1\expandafter \@firstoftwo
 \else \expandafter \@secondoftwo
 \fi
}%
\providecommand \@ifx [1]{%
 \ifx #1\expandafter \@firstoftwo
 \else \expandafter \@secondoftwo
 \fi
}%
\providecommand \natexlab [1]{#1}%
\providecommand \enquote  [1]{``#1''}%
\providecommand \bibnamefont  [1]{#1}%
\providecommand \bibfnamefont [1]{#1}%
\providecommand \citenamefont [1]{#1}%
\providecommand \href@noop [0]{\@secondoftwo}%
\providecommand \href [0]{\begingroup \@sanitize@url \@href}%
\providecommand \@href[1]{\@@startlink{#1}\@@href}%
\providecommand \@@href[1]{\endgroup#1\@@endlink}%
\providecommand \@sanitize@url [0]{\catcode `\\12\catcode `\$12\catcode
  `\&12\catcode `\#12\catcode `\^12\catcode `\_12\catcode `\%12\relax}%
\providecommand \@@startlink[1]{}%
\providecommand \@@endlink[0]{}%
\providecommand \url  [0]{\begingroup\@sanitize@url \@url }%
\providecommand \@url [1]{\endgroup\@href {#1}{\urlprefix }}%
\providecommand \urlprefix  [0]{URL }%
\providecommand \Eprint [0]{\href }%
\providecommand \doibase [0]{http://dx.doi.org/}%
\providecommand \selectlanguage [0]{\@gobble}%
\providecommand \bibinfo  [0]{\@secondoftwo}%
\providecommand \bibfield  [0]{\@secondoftwo}%
\providecommand \translation [1]{[#1]}%
\providecommand \BibitemOpen [0]{}%
\providecommand \bibitemStop [0]{}%
\providecommand \bibitemNoStop [0]{.\EOS\space}%
\providecommand \EOS [0]{\spacefactor3000\relax}%
\providecommand \BibitemShut  [1]{\csname bibitem#1\endcsname}%
\let\auto@bib@innerbib\@empty
\bibitem [{\citenamefont {Weng}\ \emph {et~al.}(2015)\citenamefont {Weng},
  \citenamefont {Fang}, \citenamefont {Fang}, \citenamefont {Bernevig},\ and\
  \citenamefont {Dai}}]{weng2015weyl}%
  \BibitemOpen
  \bibfield  {author} {\bibinfo {author} {\bibfnamefont {H.}~\bibnamefont
  {Weng}}, \bibinfo {author} {\bibfnamefont {C.}~\bibnamefont {Fang}}, \bibinfo
  {author} {\bibfnamefont {Z.}~\bibnamefont {Fang}}, \bibinfo {author}
  {\bibfnamefont {B.~A.}\ \bibnamefont {Bernevig}}, \ and\ \bibinfo {author}
  {\bibfnamefont {X.}~\bibnamefont {Dai}},\ }\href {\doibase
  10.1103/PhysRevX.5.011029} {\bibfield  {journal} {\bibinfo  {journal} {Phys.
  Rev. X}\ }\textbf {\bibinfo {volume} {5}},\ \bibinfo {pages} {011029}
  (\bibinfo {year} {2015})}\BibitemShut {NoStop}%
\bibitem [{\citenamefont {Huang}\ \emph {et~al.}(2015)\citenamefont {Huang},
  \citenamefont {Xu}, \citenamefont {Belopolski}, \citenamefont {Lee},
  \citenamefont {Chang}, \citenamefont {Wang}, \citenamefont {Alidoust},
  \citenamefont {Bian}, \citenamefont {Neupane}, \citenamefont {Zhang} \emph
  {et~al.}}]{huang2015weyl}%
  \BibitemOpen
  \bibfield  {author} {\bibinfo {author} {\bibfnamefont {S.-M.}\ \bibnamefont
  {Huang}}, \bibinfo {author} {\bibfnamefont {S.-Y.}\ \bibnamefont {Xu}},
  \bibinfo {author} {\bibfnamefont {I.}~\bibnamefont {Belopolski}}, \bibinfo
  {author} {\bibfnamefont {C.-C.}\ \bibnamefont {Lee}}, \bibinfo {author}
  {\bibfnamefont {G.}~\bibnamefont {Chang}}, \bibinfo {author} {\bibfnamefont
  {B.}~\bibnamefont {Wang}}, \bibinfo {author} {\bibfnamefont {N.}~\bibnamefont
  {Alidoust}}, \bibinfo {author} {\bibfnamefont {G.}~\bibnamefont {Bian}},
  \bibinfo {author} {\bibfnamefont {M.}~\bibnamefont {Neupane}}, \bibinfo
  {author} {\bibfnamefont {C.}~\bibnamefont {Zhang}},  \emph {et~al.},\ }\href
  {\doibase 10.1038/ncomms8373} {\bibfield  {journal} {\bibinfo  {journal}
  {Nature Comm.}\ }\textbf {\bibinfo {volume} {6}},\ \bibinfo {pages} {7373}
  (\bibinfo {year} {2015})}\BibitemShut {NoStop}%
\bibitem [{\citenamefont {Lv}\ \emph {et~al.}(2015{\natexlab{a}})\citenamefont
  {Lv}, \citenamefont {Weng}, \citenamefont {Fu}, \citenamefont {Wang},
  \citenamefont {Miao}, \citenamefont {Ma}, \citenamefont {Richard},
  \citenamefont {Huang}, \citenamefont {Zhao}, \citenamefont {Chen} \emph
  {et~al.}}]{Lv2015Experimental}%
  \BibitemOpen
  \bibfield  {author} {\bibinfo {author} {\bibfnamefont {B.~Q.}\ \bibnamefont
  {Lv}}, \bibinfo {author} {\bibfnamefont {H.~M.}\ \bibnamefont {Weng}},
  \bibinfo {author} {\bibfnamefont {B.~B.}\ \bibnamefont {Fu}}, \bibinfo
  {author} {\bibfnamefont {X.~P.}\ \bibnamefont {Wang}}, \bibinfo {author}
  {\bibfnamefont {H.}~\bibnamefont {Miao}}, \bibinfo {author} {\bibfnamefont
  {J.}~\bibnamefont {Ma}}, \bibinfo {author} {\bibfnamefont {P.}~\bibnamefont
  {Richard}}, \bibinfo {author} {\bibfnamefont {X.~C.}\ \bibnamefont {Huang}},
  \bibinfo {author} {\bibfnamefont {L.~X.}\ \bibnamefont {Zhao}}, \bibinfo
  {author} {\bibfnamefont {G.~F.}\ \bibnamefont {Chen}},  \emph {et~al.},\
  }\href {\doibase 10.1103/PhysRevX.5.031013} {\bibfield  {journal} {\bibinfo
  {journal} {Phys. Rev. X}\ }\textbf {\bibinfo {volume} {5}},\ \bibinfo {pages}
  {031013} (\bibinfo {year} {2015}{\natexlab{a}})}\BibitemShut {NoStop}%
\bibitem [{\citenamefont {Xu}\ \emph {et~al.}(2015{\natexlab{a}})\citenamefont
  {Xu}, \citenamefont {Belopolski}, \citenamefont {Alidoust}, \citenamefont
  {Neupane}, \citenamefont {Bian}, \citenamefont {Zhang}, \citenamefont
  {Sankar}, \citenamefont {Chang}, \citenamefont {Yuan}, \citenamefont {Lee}
  \emph {et~al.}}]{xu2015discovery}%
  \BibitemOpen
  \bibfield  {author} {\bibinfo {author} {\bibfnamefont {S.-Y.}\ \bibnamefont
  {Xu}}, \bibinfo {author} {\bibfnamefont {I.}~\bibnamefont {Belopolski}},
  \bibinfo {author} {\bibfnamefont {N.}~\bibnamefont {Alidoust}}, \bibinfo
  {author} {\bibfnamefont {M.}~\bibnamefont {Neupane}}, \bibinfo {author}
  {\bibfnamefont {G.}~\bibnamefont {Bian}}, \bibinfo {author} {\bibfnamefont
  {C.}~\bibnamefont {Zhang}}, \bibinfo {author} {\bibfnamefont
  {R.}~\bibnamefont {Sankar}}, \bibinfo {author} {\bibfnamefont
  {G.}~\bibnamefont {Chang}}, \bibinfo {author} {\bibfnamefont
  {Z.}~\bibnamefont {Yuan}}, \bibinfo {author} {\bibfnamefont {C.-C.}\
  \bibnamefont {Lee}},  \emph {et~al.},\ }\href {\doibase
  10.1126/science.aaa9297} {\bibfield  {journal} {\bibinfo  {journal}
  {Science}\ }\textbf {\bibinfo {volume} {349}},\ \bibinfo {pages} {613}
  (\bibinfo {year} {2015}{\natexlab{a}})}\BibitemShut {NoStop}%
\bibitem [{\citenamefont {Xu}\ \emph {et~al.}(2015{\natexlab{b}})\citenamefont
  {Xu}, \citenamefont {Alidoust}, \citenamefont {Belopolski}, \citenamefont
  {Yuan}, \citenamefont {Bian}, \citenamefont {Chang}, \citenamefont {Zheng},
  \citenamefont {Strocov}, \citenamefont {Sanchez}, \citenamefont {Chang} \emph
  {et~al.}}]{xu2015discovery2}%
  \BibitemOpen
  \bibfield  {author} {\bibinfo {author} {\bibfnamefont {S.-Y.}\ \bibnamefont
  {Xu}}, \bibinfo {author} {\bibfnamefont {N.}~\bibnamefont {Alidoust}},
  \bibinfo {author} {\bibfnamefont {I.}~\bibnamefont {Belopolski}}, \bibinfo
  {author} {\bibfnamefont {Z.}~\bibnamefont {Yuan}}, \bibinfo {author}
  {\bibfnamefont {G.}~\bibnamefont {Bian}}, \bibinfo {author} {\bibfnamefont
  {T.-R.}\ \bibnamefont {Chang}}, \bibinfo {author} {\bibfnamefont
  {H.}~\bibnamefont {Zheng}}, \bibinfo {author} {\bibfnamefont {V.~N.}\
  \bibnamefont {Strocov}}, \bibinfo {author} {\bibfnamefont {D.~S.}\
  \bibnamefont {Sanchez}}, \bibinfo {author} {\bibfnamefont {G.}~\bibnamefont
  {Chang}},  \emph {et~al.},\ }\href {\doibase 10.1038/nphys3437} {\bibfield
  {journal} {\bibinfo  {journal} {Nature Physics}\ }\textbf {\bibinfo {volume}
  {11}},\ \bibinfo {pages} {748} (\bibinfo {year}
  {2015}{\natexlab{b}})}\BibitemShut {NoStop}%
\bibitem [{\citenamefont {Yang}\ \emph {et~al.}(2015)\citenamefont {Yang},
  \citenamefont {Liu}, \citenamefont {Sun}, \citenamefont {Peng}, \citenamefont
  {Yang}, \citenamefont {Zhang}, \citenamefont {Zhou}, \citenamefont {Zhang},
  \citenamefont {Guo}, \citenamefont {Rahn} \emph {et~al.}}]{yang2015weyl}%
  \BibitemOpen
  \bibfield  {author} {\bibinfo {author} {\bibfnamefont {L.}~\bibnamefont
  {Yang}}, \bibinfo {author} {\bibfnamefont {Z.}~\bibnamefont {Liu}}, \bibinfo
  {author} {\bibfnamefont {Y.}~\bibnamefont {Sun}}, \bibinfo {author}
  {\bibfnamefont {H.}~\bibnamefont {Peng}}, \bibinfo {author} {\bibfnamefont
  {H.}~\bibnamefont {Yang}}, \bibinfo {author} {\bibfnamefont {T.}~\bibnamefont
  {Zhang}}, \bibinfo {author} {\bibfnamefont {B.}~\bibnamefont {Zhou}},
  \bibinfo {author} {\bibfnamefont {Y.}~\bibnamefont {Zhang}}, \bibinfo
  {author} {\bibfnamefont {Y.}~\bibnamefont {Guo}}, \bibinfo {author}
  {\bibfnamefont {M.}~\bibnamefont {Rahn}},  \emph {et~al.},\ }\href {\doibase
  10.1038/nphys3425} {\bibfield  {journal} {\bibinfo  {journal} {Nature
  Physics}\ }\textbf {\bibinfo {volume} {11}},\ \bibinfo {pages} {728}
  (\bibinfo {year} {2015})}\BibitemShut {NoStop}%
\bibitem [{\citenamefont {Lv}\ \emph {et~al.}(2015{\natexlab{b}})\citenamefont
  {Lv}, \citenamefont {Xu}, \citenamefont {Weng}, \citenamefont {Ma},
  \citenamefont {Richard}, \citenamefont {Huang}, \citenamefont {Zhao},
  \citenamefont {Chen}, \citenamefont {Matt}, \citenamefont {Bisti} \emph
  {et~al.}}]{lv2015observation}%
  \BibitemOpen
  \bibfield  {author} {\bibinfo {author} {\bibfnamefont {B.}~\bibnamefont
  {Lv}}, \bibinfo {author} {\bibfnamefont {N.}~\bibnamefont {Xu}}, \bibinfo
  {author} {\bibfnamefont {H.}~\bibnamefont {Weng}}, \bibinfo {author}
  {\bibfnamefont {J.}~\bibnamefont {Ma}}, \bibinfo {author} {\bibfnamefont
  {P.}~\bibnamefont {Richard}}, \bibinfo {author} {\bibfnamefont
  {X.}~\bibnamefont {Huang}}, \bibinfo {author} {\bibfnamefont
  {L.}~\bibnamefont {Zhao}}, \bibinfo {author} {\bibfnamefont {G.}~\bibnamefont
  {Chen}}, \bibinfo {author} {\bibfnamefont {C.}~\bibnamefont {Matt}}, \bibinfo
  {author} {\bibfnamefont {F.}~\bibnamefont {Bisti}},  \emph {et~al.},\ }\href
  {\doibase 10.1038/nphys3426} {\bibfield  {journal} {\bibinfo  {journal}
  {Nature Physics}\ }\textbf {\bibinfo {volume} {11}},\ \bibinfo {pages} {724}
  (\bibinfo {year} {2015}{\natexlab{b}})}\BibitemShut {NoStop}%
\bibitem [{\citenamefont {Xu}\ \emph {et~al.}(2015{\natexlab{c}})\citenamefont
  {Xu}, \citenamefont {Weng}, \citenamefont {Lv}, \citenamefont {Matt},
  \citenamefont {Park}, \citenamefont {Bisti}, \citenamefont {Strocov},
  \citenamefont {Pomjakushina}, \citenamefont {Conder}, \citenamefont {Plumb}
  \emph {et~al.}}]{xu2016observation}%
  \BibitemOpen
  \bibfield  {author} {\bibinfo {author} {\bibfnamefont {N.}~\bibnamefont
  {Xu}}, \bibinfo {author} {\bibfnamefont {H.}~\bibnamefont {Weng}}, \bibinfo
  {author} {\bibfnamefont {B.}~\bibnamefont {Lv}}, \bibinfo {author}
  {\bibfnamefont {C.}~\bibnamefont {Matt}}, \bibinfo {author} {\bibfnamefont
  {J.}~\bibnamefont {Park}}, \bibinfo {author} {\bibfnamefont {F.}~\bibnamefont
  {Bisti}}, \bibinfo {author} {\bibfnamefont {V.}~\bibnamefont {Strocov}},
  \bibinfo {author} {\bibfnamefont {E.}~\bibnamefont {Pomjakushina}}, \bibinfo
  {author} {\bibfnamefont {K.}~\bibnamefont {Conder}}, \bibinfo {author}
  {\bibfnamefont {N.}~\bibnamefont {Plumb}},  \emph {et~al.},\ }\href {\doibase
  10.1038/ncomms11006} {\bibfield  {journal} {\bibinfo  {journal} {Nature
  Comm.}\ }\textbf {\bibinfo {volume} {7}},\ \bibinfo {pages} {11006} (\bibinfo
  {year} {2015}{\natexlab{c}})}\BibitemShut {NoStop}%
\bibitem [{\citenamefont {Shekhar}\ \emph {et~al.}(2015)\citenamefont
  {Shekhar}, \citenamefont {Nayak}, \citenamefont {Sun}, \citenamefont
  {Schmidt}, \citenamefont {Nicklas}, \citenamefont {Leermakers}, \citenamefont
  {Zeitler}, \citenamefont {Schnelle}, \citenamefont {Grin}, \citenamefont
  {Felser} \emph {et~al.}}]{shekhar2015extremely}%
  \BibitemOpen
  \bibfield  {author} {\bibinfo {author} {\bibfnamefont {C.}~\bibnamefont
  {Shekhar}}, \bibinfo {author} {\bibfnamefont {A.~K.}\ \bibnamefont {Nayak}},
  \bibinfo {author} {\bibfnamefont {Y.}~\bibnamefont {Sun}}, \bibinfo {author}
  {\bibfnamefont {M.}~\bibnamefont {Schmidt}}, \bibinfo {author} {\bibfnamefont
  {M.}~\bibnamefont {Nicklas}}, \bibinfo {author} {\bibfnamefont
  {I.}~\bibnamefont {Leermakers}}, \bibinfo {author} {\bibfnamefont
  {U.}~\bibnamefont {Zeitler}}, \bibinfo {author} {\bibfnamefont
  {W.}~\bibnamefont {Schnelle}}, \bibinfo {author} {\bibfnamefont
  {J.}~\bibnamefont {Grin}}, \bibinfo {author} {\bibfnamefont {C.}~\bibnamefont
  {Felser}},  \emph {et~al.},\ }\href {\doibase 10.1038/nphys3372} {\bibfield
  {journal} {\bibinfo  {journal} {Nature Physics}\ }\textbf {\bibinfo {volume}
  {11}},\ \bibinfo {pages} {645} (\bibinfo {year} {2015})}\BibitemShut
  {NoStop}%
\bibitem [{\citenamefont {Shi}\ and\ \citenamefont {Song}(2017)}]{Shi2017}%
  \BibitemOpen
  \bibfield  {author} {\bibinfo {author} {\bibfnamefont {L.-k.}\ \bibnamefont
  {Shi}}\ and\ \bibinfo {author} {\bibfnamefont {J.}~\bibnamefont {Song}},\
  }\href {https://arxiv.org/abs/1705.01566} {\bibfield  {journal} {\bibinfo
  {journal} {Preprint, arXiv:1705.01566}\ } (\bibinfo {year}
  {2017})}\BibitemShut {NoStop}%
\bibitem [{\citenamefont {Baum}\ \emph {et~al.}(2015)\citenamefont {Baum},
  \citenamefont {Berg}, \citenamefont {Parameswaran},\ and\ \citenamefont
  {Stern}}]{Baum2015}%
  \BibitemOpen
  \bibfield  {author} {\bibinfo {author} {\bibfnamefont {Y.}~\bibnamefont
  {Baum}}, \bibinfo {author} {\bibfnamefont {E.}~\bibnamefont {Berg}}, \bibinfo
  {author} {\bibfnamefont {S.~A.}\ \bibnamefont {Parameswaran}}, \ and\
  \bibinfo {author} {\bibfnamefont {A.}~\bibnamefont {Stern}},\ }\href
  {\doibase 10.1103/PhysRevX.5.041046} {\bibfield  {journal} {\bibinfo
  {journal} {Phys. Rev. X}\ }\textbf {\bibinfo {volume} {5}},\ \bibinfo {pages}
  {041046} (\bibinfo {year} {2015})}\BibitemShut {NoStop}%
\bibitem [{\citenamefont {Bera}\ \emph {et~al.}(2016)\citenamefont {Bera},
  \citenamefont {Sau},\ and\ \citenamefont {Roy}}]{bera2016dirty}%
  \BibitemOpen
  \bibfield  {author} {\bibinfo {author} {\bibfnamefont {S.}~\bibnamefont
  {Bera}}, \bibinfo {author} {\bibfnamefont {J.~D.}\ \bibnamefont {Sau}}, \
  and\ \bibinfo {author} {\bibfnamefont {B.}~\bibnamefont {Roy}},\ }\href
  {\doibase 10.1103/PhysRevB.93.201302} {\bibfield  {journal} {\bibinfo
  {journal} {Phys. Rev. B}\ }\textbf {\bibinfo {volume} {93}},\ \bibinfo
  {pages} {201302} (\bibinfo {year} {2016})}\BibitemShut {NoStop}%
\bibitem [{\citenamefont {Wan}\ \emph {et~al.}(2011)\citenamefont {Wan},
  \citenamefont {Turner}, \citenamefont {Vishwanath},\ and\ \citenamefont
  {Savrasov}}]{wan2011topological}%
  \BibitemOpen
  \bibfield  {author} {\bibinfo {author} {\bibfnamefont {X.}~\bibnamefont
  {Wan}}, \bibinfo {author} {\bibfnamefont {A.~M.}\ \bibnamefont {Turner}},
  \bibinfo {author} {\bibfnamefont {A.}~\bibnamefont {Vishwanath}}, \ and\
  \bibinfo {author} {\bibfnamefont {S.~Y.}\ \bibnamefont {Savrasov}},\ }\href
  {\doibase 10.1103/PhysRevB.83.205101} {\bibfield  {journal} {\bibinfo
  {journal} {Phys. Rev. B}\ }\textbf {\bibinfo {volume} {83}},\ \bibinfo
  {pages} {205101} (\bibinfo {year} {2011})}\BibitemShut {NoStop}%
\bibitem [{\citenamefont {Anderson}(1958)}]{Anderson1958}%
  \BibitemOpen
  \bibfield  {author} {\bibinfo {author} {\bibfnamefont {P.~W.}\ \bibnamefont
  {Anderson}},\ }\href {\doibase 10.1103/PhysRev.109.1492} {\bibfield
  {journal} {\bibinfo  {journal} {Phys. Rev.}\ }\textbf {\bibinfo {volume}
  {109}},\ \bibinfo {pages} {1492} (\bibinfo {year} {1958})}\BibitemShut
  {NoStop}%
\bibitem [{\citenamefont {Abrahams}\ \emph {et~al.}(1979)\citenamefont
  {Abrahams}, \citenamefont {Anderson}, \citenamefont {Licciardello},\ and\
  \citenamefont {Ramakrishnan}}]{Abrahams1979}%
  \BibitemOpen
  \bibfield  {author} {\bibinfo {author} {\bibfnamefont {E.}~\bibnamefont
  {Abrahams}}, \bibinfo {author} {\bibfnamefont {P.~W.}\ \bibnamefont
  {Anderson}}, \bibinfo {author} {\bibfnamefont {D.~C.}\ \bibnamefont
  {Licciardello}}, \ and\ \bibinfo {author} {\bibfnamefont {T.~V.}\
  \bibnamefont {Ramakrishnan}},\ }\href {\doibase 10.1103/PhysRevLett.42.673}
  {\bibfield  {journal} {\bibinfo  {journal} {Phys. Rev. Lett.}\ }\textbf
  {\bibinfo {volume} {42}},\ \bibinfo {pages} {673} (\bibinfo {year}
  {1979})}\BibitemShut {NoStop}%
\bibitem [{\citenamefont {Bardarson}\ \emph {et~al.}(2007)\citenamefont
  {Bardarson}, \citenamefont {Tworzyd\l{}o}, \citenamefont {Brouwer},\ and\
  \citenamefont {Beenakker}}]{Bardarson2007}%
  \BibitemOpen
  \bibfield  {author} {\bibinfo {author} {\bibfnamefont {J.~H.}\ \bibnamefont
  {Bardarson}}, \bibinfo {author} {\bibfnamefont {J.}~\bibnamefont
  {Tworzyd\l{}o}}, \bibinfo {author} {\bibfnamefont {P.~W.}\ \bibnamefont
  {Brouwer}}, \ and\ \bibinfo {author} {\bibfnamefont {C.~W.~J.}\ \bibnamefont
  {Beenakker}},\ }\href {\doibase 10.1103/PhysRevLett.99.106801} {\bibfield
  {journal} {\bibinfo  {journal} {Phys. Rev. Lett.}\ }\textbf {\bibinfo
  {volume} {99}},\ \bibinfo {pages} {106801} (\bibinfo {year}
  {2007})}\BibitemShut {NoStop}%
\bibitem [{\citenamefont {Altland}\ and\ \citenamefont
  {Bagrets}(2016)}]{Altland2016}%
  \BibitemOpen
  \bibfield  {author} {\bibinfo {author} {\bibfnamefont {A.}~\bibnamefont
  {Altland}}\ and\ \bibinfo {author} {\bibfnamefont {D.}~\bibnamefont
  {Bagrets}},\ }\href {\doibase 10.1103/PhysRevB.93.075113} {\bibfield
  {journal} {\bibinfo  {journal} {Phys. Rev. B}\ }\textbf {\bibinfo {volume}
  {93}},\ \bibinfo {pages} {075113} (\bibinfo {year} {2016})}\BibitemShut
  {NoStop}%
\bibitem [{\citenamefont {Xiong}\ \emph {et~al.}(2015)\citenamefont {Xiong},
  \citenamefont {Kushwaha}, \citenamefont {Liang}, \citenamefont {Krizan},
  \citenamefont {Wang}, \citenamefont {Cava},\ and\ \citenamefont
  {Ong}}]{Xiong2015}%
  \BibitemOpen
  \bibfield  {author} {\bibinfo {author} {\bibfnamefont {J.}~\bibnamefont
  {Xiong}}, \bibinfo {author} {\bibfnamefont {S.~K.}\ \bibnamefont {Kushwaha}},
  \bibinfo {author} {\bibfnamefont {T.}~\bibnamefont {Liang}}, \bibinfo
  {author} {\bibfnamefont {J.~W.}\ \bibnamefont {Krizan}}, \bibinfo {author}
  {\bibfnamefont {W.}~\bibnamefont {Wang}}, \bibinfo {author} {\bibfnamefont
  {R.~J.}\ \bibnamefont {Cava}}, \ and\ \bibinfo {author} {\bibfnamefont
  {N.~P.}\ \bibnamefont {Ong}},\ }\href {https://arxiv.org/abs/1503.08179}
  {\bibfield  {journal} {\bibinfo  {journal} {Preprint, arXiv:1503.08179}\ }
  (\bibinfo {year} {2015})}\BibitemShut {NoStop}%
\bibitem [{\citenamefont {Fradkin}(1986)}]{Fradkin1986}%
  \BibitemOpen
  \bibfield  {author} {\bibinfo {author} {\bibfnamefont {E.}~\bibnamefont
  {Fradkin}},\ }\href {\doibase 10.1103/PhysRevB.33.3263} {\bibfield  {journal}
  {\bibinfo  {journal} {Phys. Rev. B}\ }\textbf {\bibinfo {volume} {33}},\
  \bibinfo {pages} {3263} (\bibinfo {year} {1986})}\BibitemShut {NoStop}%
\bibitem [{\citenamefont {Ominato}\ and\ \citenamefont
  {Koshino}(2014)}]{Ominato2014}%
  \BibitemOpen
  \bibfield  {author} {\bibinfo {author} {\bibfnamefont {Y.}~\bibnamefont
  {Ominato}}\ and\ \bibinfo {author} {\bibfnamefont {M.}~\bibnamefont
  {Koshino}},\ }\href {\doibase 10.1103/PhysRevB.89.054202} {\bibfield
  {journal} {\bibinfo  {journal} {Phys. Rev. B}\ }\textbf {\bibinfo {volume}
  {89}},\ \bibinfo {pages} {054202} (\bibinfo {year} {2014})}\BibitemShut
  {NoStop}%
\bibitem [{\citenamefont {Sbierski}\ \emph {et~al.}(2014)\citenamefont
  {Sbierski}, \citenamefont {Pohl}, \citenamefont {Bergholtz},\ and\
  \citenamefont {Brouwer}}]{Sbierski2014}%
  \BibitemOpen
  \bibfield  {author} {\bibinfo {author} {\bibfnamefont {B.}~\bibnamefont
  {Sbierski}}, \bibinfo {author} {\bibfnamefont {G.}~\bibnamefont {Pohl}},
  \bibinfo {author} {\bibfnamefont {E.~J.}\ \bibnamefont {Bergholtz}}, \ and\
  \bibinfo {author} {\bibfnamefont {P.~W.}\ \bibnamefont {Brouwer}},\ }\href
  {\doibase 10.1103/PhysRevLett.113.026602} {\bibfield  {journal} {\bibinfo
  {journal} {Phys. Rev. Lett.}\ }\textbf {\bibinfo {volume} {113}},\ \bibinfo
  {pages} {026602} (\bibinfo {year} {2014})}\BibitemShut {NoStop}%
\bibitem [{\citenamefont {Syzranov}\ \emph {et~al.}(2015)\citenamefont
  {Syzranov}, \citenamefont {Radzihovsky},\ and\ \citenamefont
  {Gurarie}}]{Syzranov2015}%
  \BibitemOpen
  \bibfield  {author} {\bibinfo {author} {\bibfnamefont {S.~V.}\ \bibnamefont
  {Syzranov}}, \bibinfo {author} {\bibfnamefont {L.}~\bibnamefont
  {Radzihovsky}}, \ and\ \bibinfo {author} {\bibfnamefont {V.}~\bibnamefont
  {Gurarie}},\ }\href {\doibase 10.1103/PhysRevLett.114.166601} {\bibfield
  {journal} {\bibinfo  {journal} {Phys. Rev. Lett.}\ }\textbf {\bibinfo
  {volume} {114}},\ \bibinfo {pages} {166601} (\bibinfo {year}
  {2015})}\BibitemShut {NoStop}%
\bibitem [{\citenamefont {Pixley}\ \emph {et~al.}(2015)\citenamefont {Pixley},
  \citenamefont {Goswami},\ and\ \citenamefont {Das~Sarma}}]{Pixley2015}%
  \BibitemOpen
  \bibfield  {author} {\bibinfo {author} {\bibfnamefont {J.~H.}\ \bibnamefont
  {Pixley}}, \bibinfo {author} {\bibfnamefont {P.}~\bibnamefont {Goswami}}, \
  and\ \bibinfo {author} {\bibfnamefont {S.}~\bibnamefont {Das~Sarma}},\ }\href
  {\doibase 10.1103/PhysRevLett.115.076601} {\bibfield  {journal} {\bibinfo
  {journal} {Phys. Rev. Lett.}\ }\textbf {\bibinfo {volume} {115}},\ \bibinfo
  {pages} {076601} (\bibinfo {year} {2015})}\BibitemShut {NoStop}%
\bibitem [{\citenamefont {Pixley}\ \emph
  {et~al.}(2016{\natexlab{a}})\citenamefont {Pixley}, \citenamefont {Goswami},\
  and\ \citenamefont {Das~Sarma}}]{Pixley2016A}%
  \BibitemOpen
  \bibfield  {author} {\bibinfo {author} {\bibfnamefont {J.~H.}\ \bibnamefont
  {Pixley}}, \bibinfo {author} {\bibfnamefont {P.}~\bibnamefont {Goswami}}, \
  and\ \bibinfo {author} {\bibfnamefont {S.}~\bibnamefont {Das~Sarma}},\ }\href
  {\doibase 10.1103/PhysRevB.93.085103} {\bibfield  {journal} {\bibinfo
  {journal} {Phys. Rev. B}\ }\textbf {\bibinfo {volume} {93}},\ \bibinfo
  {pages} {085103} (\bibinfo {year} {2016}{\natexlab{a}})}\BibitemShut
  {NoStop}%
\bibitem [{\citenamefont {Pixley}\ \emph
  {et~al.}(2016{\natexlab{b}})\citenamefont {Pixley}, \citenamefont {Huse},\
  and\ \citenamefont {Das~Sarma}}]{Pixley2016}%
  \BibitemOpen
  \bibfield  {author} {\bibinfo {author} {\bibfnamefont {J.~H.}\ \bibnamefont
  {Pixley}}, \bibinfo {author} {\bibfnamefont {D.~A.}\ \bibnamefont {Huse}}, \
  and\ \bibinfo {author} {\bibfnamefont {S.}~\bibnamefont {Das~Sarma}},\ }\href
  {\doibase 10.1103/PhysRevX.6.021042} {\bibfield  {journal} {\bibinfo
  {journal} {Phys. Rev. X}\ }\textbf {\bibinfo {volume} {6}},\ \bibinfo {pages}
  {021042} (\bibinfo {year} {2016}{\natexlab{b}})}\BibitemShut {NoStop}%
\bibitem [{\citenamefont {Rodionov}\ \emph {et~al.}(2015)\citenamefont
  {Rodionov}, \citenamefont {Kugel},\ and\ \citenamefont
  {Nori}}]{Rodionov2015}%
  \BibitemOpen
  \bibfield  {author} {\bibinfo {author} {\bibfnamefont {Y.~I.}\ \bibnamefont
  {Rodionov}}, \bibinfo {author} {\bibfnamefont {K.~I.}\ \bibnamefont {Kugel}},
  \ and\ \bibinfo {author} {\bibfnamefont {F.}~\bibnamefont {Nori}},\ }\href
  {\doibase 10.1103/PhysRevB.92.195117} {\bibfield  {journal} {\bibinfo
  {journal} {Phys. Rev. B}\ }\textbf {\bibinfo {volume} {92}},\ \bibinfo
  {pages} {195117} (\bibinfo {year} {2015})}\BibitemShut {NoStop}%
\bibitem [{\citenamefont {Trescher}\ \emph {et~al.}(2015)\citenamefont
  {Trescher}, \citenamefont {Sbierski}, \citenamefont {Brouwer},\ and\
  \citenamefont {Bergholtz}}]{Trescher2015}%
  \BibitemOpen
  \bibfield  {author} {\bibinfo {author} {\bibfnamefont {M.}~\bibnamefont
  {Trescher}}, \bibinfo {author} {\bibfnamefont {B.}~\bibnamefont {Sbierski}},
  \bibinfo {author} {\bibfnamefont {P.~W.}\ \bibnamefont {Brouwer}}, \ and\
  \bibinfo {author} {\bibfnamefont {E.~J.}\ \bibnamefont {Bergholtz}},\ }\href
  {\doibase 10.1103/PhysRevB.91.115135} {\bibfield  {journal} {\bibinfo
  {journal} {Phys. Rev. B}\ }\textbf {\bibinfo {volume} {91}},\ \bibinfo
  {pages} {115135} (\bibinfo {year} {2015})}\BibitemShut {NoStop}%
\bibitem [{\citenamefont {Soluyanov}\ \emph {et~al.}(2015)\citenamefont
  {Soluyanov}, \citenamefont {Gresch}, \citenamefont {Wang}, \citenamefont
  {Wu}, \citenamefont {Troyer}, \citenamefont {Dai},\ and\ \citenamefont
  {Bernevig}}]{soluyanov2015typeii}%
  \BibitemOpen
  \bibfield  {author} {\bibinfo {author} {\bibfnamefont {A.~A.}\ \bibnamefont
  {Soluyanov}}, \bibinfo {author} {\bibfnamefont {D.}~\bibnamefont {Gresch}},
  \bibinfo {author} {\bibfnamefont {Z.}~\bibnamefont {Wang}}, \bibinfo {author}
  {\bibfnamefont {Q.}~\bibnamefont {Wu}}, \bibinfo {author} {\bibfnamefont
  {M.}~\bibnamefont {Troyer}}, \bibinfo {author} {\bibfnamefont
  {X.}~\bibnamefont {Dai}}, \ and\ \bibinfo {author} {\bibfnamefont {B.~A.}\
  \bibnamefont {Bernevig}},\ }\href {\doibase 10.1038/nature15768} {\bibfield
  {journal} {\bibinfo  {journal} {Nature}\ }\textbf {\bibinfo {volume} {527}},\
  \bibinfo {pages} {495} (\bibinfo {year} {2015})}\BibitemShut {NoStop}%
\bibitem [{\citenamefont {Bergholtz}\ \emph {et~al.}(2015)\citenamefont
  {Bergholtz}, \citenamefont {Liu}, \citenamefont {Trescher}, \citenamefont
  {Moessner},\ and\ \citenamefont {Udagawa}}]{Bergholtz2015}%
  \BibitemOpen
  \bibfield  {author} {\bibinfo {author} {\bibfnamefont {E.~J.}\ \bibnamefont
  {Bergholtz}}, \bibinfo {author} {\bibfnamefont {Z.}~\bibnamefont {Liu}},
  \bibinfo {author} {\bibfnamefont {M.}~\bibnamefont {Trescher}}, \bibinfo
  {author} {\bibfnamefont {R.}~\bibnamefont {Moessner}}, \ and\ \bibinfo
  {author} {\bibfnamefont {M.}~\bibnamefont {Udagawa}},\ }\href {\doibase
  10.1103/PhysRevLett.114.016806} {\bibfield  {journal} {\bibinfo  {journal}
  {Phys. Rev. Lett.}\ }\textbf {\bibinfo {volume} {114}},\ \bibinfo {pages}
  {016806} (\bibinfo {year} {2015})}\BibitemShut {NoStop}%
\bibitem [{\citenamefont {Koepernik}\ \emph {et~al.}(2016)\citenamefont
  {Koepernik}, \citenamefont {Kasinathan}, \citenamefont {Efremov},
  \citenamefont {Khim}, \citenamefont {Borisenko}, \citenamefont {B\"uchner},\
  and\ \citenamefont {van~den Brink}}]{Koepernik2016}%
  \BibitemOpen
  \bibfield  {author} {\bibinfo {author} {\bibfnamefont {K.}~\bibnamefont
  {Koepernik}}, \bibinfo {author} {\bibfnamefont {D.}~\bibnamefont
  {Kasinathan}}, \bibinfo {author} {\bibfnamefont {D.~V.}\ \bibnamefont
  {Efremov}}, \bibinfo {author} {\bibfnamefont {S.}~\bibnamefont {Khim}},
  \bibinfo {author} {\bibfnamefont {S.}~\bibnamefont {Borisenko}}, \bibinfo
  {author} {\bibfnamefont {B.}~\bibnamefont {B\"uchner}}, \ and\ \bibinfo
  {author} {\bibfnamefont {J.}~\bibnamefont {van~den Brink}},\ }\href {\doibase
  10.1103/PhysRevB.93.201101} {\bibfield  {journal} {\bibinfo  {journal} {Phys.
  Rev. B}\ }\textbf {\bibinfo {volume} {93}},\ \bibinfo {pages} {201101}
  (\bibinfo {year} {2016})}\BibitemShut {NoStop}%
\bibitem [{\citenamefont {Huang}\ \emph {et~al.}(2016)\citenamefont {Huang},
  \citenamefont {McCormick}, \citenamefont {Ochi}, \citenamefont {Zhao},
  \citenamefont {Suzuki}, \citenamefont {Arita}, \citenamefont {Wu},
  \citenamefont {Mou}, \citenamefont {Cao}, \citenamefont {Yan} \emph
  {et~al.}}]{Huang2016}%
  \BibitemOpen
  \bibfield  {author} {\bibinfo {author} {\bibfnamefont {L.}~\bibnamefont
  {Huang}}, \bibinfo {author} {\bibfnamefont {T.~M.}\ \bibnamefont
  {McCormick}}, \bibinfo {author} {\bibfnamefont {M.}~\bibnamefont {Ochi}},
  \bibinfo {author} {\bibfnamefont {Z.}~\bibnamefont {Zhao}}, \bibinfo {author}
  {\bibfnamefont {M.-T.}\ \bibnamefont {Suzuki}}, \bibinfo {author}
  {\bibfnamefont {R.}~\bibnamefont {Arita}}, \bibinfo {author} {\bibfnamefont
  {Y.}~\bibnamefont {Wu}}, \bibinfo {author} {\bibfnamefont {D.}~\bibnamefont
  {Mou}}, \bibinfo {author} {\bibfnamefont {H.}~\bibnamefont {Cao}}, \bibinfo
  {author} {\bibfnamefont {J.}~\bibnamefont {Yan}},  \emph {et~al.},\ }\href
  {\doibase 10.1038/nmat4685} {\bibfield  {journal} {\bibinfo  {journal}
  {Nature Mater.}\ }\textbf {\bibinfo {volume} {15}},\ \bibinfo {pages}
  {1155—} (\bibinfo {year} {2016})}\BibitemShut {NoStop}%
\bibitem [{\citenamefont {Belopolski}\ \emph {et~al.}(2016)\citenamefont
  {Belopolski}, \citenamefont {Sanchez}, \citenamefont {Ishida}, \citenamefont
  {Pan}, \citenamefont {Yu}, \citenamefont {Xu}, \citenamefont {Chang},
  \citenamefont {Chang}, \citenamefont {Zheng}, \citenamefont {Alidoust} \emph
  {et~al.}}]{Belopolski2016}%
  \BibitemOpen
  \bibfield  {author} {\bibinfo {author} {\bibfnamefont {I.}~\bibnamefont
  {Belopolski}}, \bibinfo {author} {\bibfnamefont {D.~S.}\ \bibnamefont
  {Sanchez}}, \bibinfo {author} {\bibfnamefont {Y.}~\bibnamefont {Ishida}},
  \bibinfo {author} {\bibfnamefont {X.}~\bibnamefont {Pan}}, \bibinfo {author}
  {\bibfnamefont {P.}~\bibnamefont {Yu}}, \bibinfo {author} {\bibfnamefont
  {S.-Y.}\ \bibnamefont {Xu}}, \bibinfo {author} {\bibfnamefont
  {G.}~\bibnamefont {Chang}}, \bibinfo {author} {\bibfnamefont {T.-R.}\
  \bibnamefont {Chang}}, \bibinfo {author} {\bibfnamefont {H.}~\bibnamefont
  {Zheng}}, \bibinfo {author} {\bibfnamefont {N.}~\bibnamefont {Alidoust}},
  \emph {et~al.},\ }\href {\doibase 10.1038/ncomms13643} {\bibfield  {journal}
  {\bibinfo  {journal} {Nature Comm.}\ }\textbf {\bibinfo {volume} {7}},\
  \bibinfo {pages} {13643} (\bibinfo {year} {2016})}\BibitemShut {NoStop}%
\bibitem [{\citenamefont {Zyuzin}\ and\ \citenamefont
  {Tiwari}(2016)}]{Zyuzin2016}%
  \BibitemOpen
  \bibfield  {author} {\bibinfo {author} {\bibfnamefont {A.~A.}\ \bibnamefont
  {Zyuzin}}\ and\ \bibinfo {author} {\bibfnamefont {R.~P.}\ \bibnamefont
  {Tiwari}},\ }\href {\doibase 10.1134/S002136401611014X} {\bibfield  {journal}
  {\bibinfo  {journal} {JETP Letters}\ }\textbf {\bibinfo {volume} {103}},\
  \bibinfo {pages} {717} (\bibinfo {year} {2016})}\BibitemShut {NoStop}%
\bibitem [{\citenamefont {O'Brien}\ \emph {et~al.}(2016)\citenamefont
  {O'Brien}, \citenamefont {Diez},\ and\ \citenamefont
  {Beenakker}}]{OBrien2016}%
  \BibitemOpen
  \bibfield  {author} {\bibinfo {author} {\bibfnamefont {T.~E.}\ \bibnamefont
  {O'Brien}}, \bibinfo {author} {\bibfnamefont {M.}~\bibnamefont {Diez}}, \
  and\ \bibinfo {author} {\bibfnamefont {C.~W.~J.}\ \bibnamefont {Beenakker}},\
  }\href {\doibase 10.1103/PhysRevLett.116.236401} {\bibfield  {journal}
  {\bibinfo  {journal} {Phys. Rev. Lett.}\ }\textbf {\bibinfo {volume} {116}},\
  \bibinfo {pages} {236401} (\bibinfo {year} {2016})}\BibitemShut {NoStop}%
\bibitem [{\citenamefont {Yu}\ \emph {et~al.}(2016)\citenamefont {Yu},
  \citenamefont {Yao},\ and\ \citenamefont {Yang}}]{Yu2016}%
  \BibitemOpen
  \bibfield  {author} {\bibinfo {author} {\bibfnamefont {Z.-M.}\ \bibnamefont
  {Yu}}, \bibinfo {author} {\bibfnamefont {Y.}~\bibnamefont {Yao}}, \ and\
  \bibinfo {author} {\bibfnamefont {S.~A.}\ \bibnamefont {Yang}},\ }\href
  {\doibase 10.1103/PhysRevLett.117.077202} {\bibfield  {journal} {\bibinfo
  {journal} {Phys. Rev. Lett.}\ }\textbf {\bibinfo {volume} {117}},\ \bibinfo
  {pages} {077202} (\bibinfo {year} {2016})}\BibitemShut {NoStop}%
\bibitem [{\citenamefont {Trescher}\ \emph {et~al.}(2017)\citenamefont
  {Trescher}, \citenamefont {Sbierski}, \citenamefont {Brouwer},\ and\
  \citenamefont {Bergholtz}}]{Trescher2017}%
  \BibitemOpen
  \bibfield  {author} {\bibinfo {author} {\bibfnamefont {M.}~\bibnamefont
  {Trescher}}, \bibinfo {author} {\bibfnamefont {B.}~\bibnamefont {Sbierski}},
  \bibinfo {author} {\bibfnamefont {P.~W.}\ \bibnamefont {Brouwer}}, \ and\
  \bibinfo {author} {\bibfnamefont {E.~J.}\ \bibnamefont {Bergholtz}},\ }\href
  {\doibase 10.1103/PhysRevB.95.045139} {\bibfield  {journal} {\bibinfo
  {journal} {Phys. Rev. B}\ }\textbf {\bibinfo {volume} {95}},\ \bibinfo
  {pages} {045139} (\bibinfo {year} {2017})}\BibitemShut {NoStop}%
\bibitem [{\citenamefont {Goswami}\ and\ \citenamefont
  {Chakravarty}(2011)}]{Goswami2011}%
  \BibitemOpen
  \bibfield  {author} {\bibinfo {author} {\bibfnamefont {P.}~\bibnamefont
  {Goswami}}\ and\ \bibinfo {author} {\bibfnamefont {S.}~\bibnamefont
  {Chakravarty}},\ }\href {\doibase 10.1103/PhysRevLett.107.196803} {\bibfield
  {journal} {\bibinfo  {journal} {Phys. Rev. Lett.}\ }\textbf {\bibinfo
  {volume} {107}},\ \bibinfo {pages} {196803} (\bibinfo {year}
  {2011})}\BibitemShut {NoStop}%
\bibitem [{\citenamefont {Roy}\ \emph {et~al.}(2016)\citenamefont {Roy},
  \citenamefont {Slager},\ and\ \citenamefont {Juricic}}]{Roy2016}%
  \BibitemOpen
  \bibfield  {author} {\bibinfo {author} {\bibfnamefont {B.}~\bibnamefont
  {Roy}}, \bibinfo {author} {\bibfnamefont {R.-J.}\ \bibnamefont {Slager}}, \
  and\ \bibinfo {author} {\bibfnamefont {V.}~\bibnamefont {Juricic}},\ }\href
  {https://arxiv.org/abs/1610.08973} {\bibfield  {journal} {\bibinfo  {journal}
  {Preprint, arXiv:1610.08973}\ } (\bibinfo {year} {2016})}\BibitemShut
  {NoStop}%
\bibitem [{\citenamefont {Roy}\ and\ \citenamefont
  {Das~Sarma}(2014)}]{Roy2014}%
  \BibitemOpen
  \bibfield  {author} {\bibinfo {author} {\bibfnamefont {B.}~\bibnamefont
  {Roy}}\ and\ \bibinfo {author} {\bibfnamefont {S.}~\bibnamefont
  {Das~Sarma}},\ }\href {\doibase 10.1103/PhysRevB.90.241112} {\bibfield
  {journal} {\bibinfo  {journal} {Phys. Rev. B}\ }\textbf {\bibinfo {volume}
  {90}},\ \bibinfo {pages} {241112} (\bibinfo {year} {2014})}\BibitemShut
  {NoStop}%
\bibitem [{\citenamefont {Detassis}\ \emph {et~al.}(2017)\citenamefont
  {Detassis}, \citenamefont {Fritz},\ and\ \citenamefont
  {Gubrinskas}}]{Detassis2017}%
  \BibitemOpen
  \bibfield  {author} {\bibinfo {author} {\bibfnamefont {F.}~\bibnamefont
  {Detassis}}, \bibinfo {author} {\bibfnamefont {L.}~\bibnamefont {Fritz}}, \
  and\ \bibinfo {author} {\bibfnamefont {S.}~\bibnamefont {Gubrinskas}},\
  }\href {https://arxiv.org/abs/1703.02425} {\bibfield  {journal} {\bibinfo
  {journal} {Preprint, arXiv:1703.02425}\ } (\bibinfo {year}
  {2017})}\BibitemShut {NoStop}%
\end{thebibliography}
\end{document}